\documentclass[graphicx,11pt,preprint]{aastex}

\begin{document}

\title{Challenges in Planet Formation} 

\author{\textbf{Alessandro Morbidelli$^{(1)}$, Sean N. Raymond$^{(2,3)}$}\\  
(1) Laboratoire Lagrange, UMR7293, Universit\'e de Nice Sophia-Antipolis,
  CNRS, Observatoire de la C\^ote d'Azur. Boulevard de l'Observatoire,
  06304 Nice Cedex 4, France. (Email: morby@oca.eu / Fax:
  +33-4-92003118) \\
(2) CNRS, Laboratoire d'Astrophysique de Bordeaux, UMR 5804, F-33270, Floirac, France. (Email: rayray.sean@gmail.com)\\
(3) Univ. Bordeaux, Laboratoire d'Astrophysique de Bordeaux, UMR 5804, F-33270, Floirac, France. \\ 
} 



\begin{abstract}
Over the past two decades, large strides have been made in the field of planet formation. Yet fundamental questions remain. Here we review our state of understanding of five fundamental bottlenecks in planet formation. These are: 1) the structure and evolution of protoplanetary disks; 2) the growth of the first planetesimals; 3) orbital migration driven by interactions between proto-planets and gaseous disk; 4) the origin of the Solar System's orbital architecture; and 5) the relationship between observed super-Earths and our own terrestrial planets. Given our lack of understanding of these issues, even the most successful formation models remain on shaky ground.  
\vskip 15pt\hbox{}
\end{abstract}

\section{Introduction}

The origin of planets is a vast, complex and still quite mysterious subject. Despite decades of space exploration, ground based observations and detailed analyses of meteorites and cometary grains (the only space samples available in our laboratories) it is still not clear how the planets of the Solar System formed. The discovery of extrasolar planets has added confusion to the problem, bringing to light evidence that planetary systems are very diverse, that our Solar System is not a typical case and that categories of planets that don't exist in our system are common elsewhere (e.g. the super-Earth planets). 

There are several recent reviews on planet formation. Johansen et al. (2014) focused on planetesimal formation, Morbidelli et al. (2012) and Raymond et al. (2014) focused on terrestrial planets, and Helled et al. (2014) focused on giant planets. These reviews remain up to date. Therefore, the goal of this chapter is to focus instead on open problems and the main issues of debate. 

Accordingly, here we identify and discuss what we (the authors of this review) consider the top 5 bottlenecks in the field of planet formation. These are:
\begin{enumerate}
\item What is the structure of protoplanetary disks?  (Section 2) 
\item How did the first planetesimals form?  (Section 3) 
\item How do planets migrate?  (Section 4) 
\item What is the origin of the {\it tri-modal} structure of the Solar System? (Section 5) 
\item What is the relationship between terrestrial planets and super-Earths? (Section 6) 
\end{enumerate}

These issues cover a range of size scales and time scales, and illustrate the upcoming challenges in planet formation.  Our choice of the ``top 5'' is admittedly biased.  Nonetheless, these issues are clear and present obstacles to our understanding of planet formation.  We conclude this review in Section 7 with a discussion.


\section{What is the structure of protoplanetary disks?}
\label{disk}

Circumstellar disks of gas and dust represent the cradles of planet formation. The problem is to understand the mechanism by which protoplanetary disks evolve and transport mass to the central star (see, e.g., Armitage 2011). Depending on which mechanism is dominant, the resulting global disk structures may differ significantly (see Figure 1), with profound effects on planet formation and migration. 

It has long been thought that transport of angular momentum in the disk is due to turbulent viscosity (Shakura and Sunayev, 1973) and that the magneto-rotational instability is the main source of turbulence ({ Balbus and Hawley, 1991}). With simple scaling arguments on the size of the vortices and their rotation period { (or, equivalently, assuming that the stress tensor scales linearly with the local gas pressure)}, Shakura and Sunayev (1973) modeled the viscosity $\nu$ as proportional to the square of disk's scale height $H$ and to rotation frequency $\Omega$:  $\nu=\alpha H^2\Omega$, where $\alpha$ is the proportionality coefficient. The assumption that $\alpha$ is the same everywhere leads to the so-called $\alpha$-disk model ({ Lyndel-Bell and Pringle, 1974; Pringle, 1981; Balbus and Papaloizou, 1999}). The structure of these disks basically depends only on three parameters: the accretion rate of gas onto the central star, the value of $\alpha$ and the dust/gas ratio, the latter governing the disk's opacity. { The disk evolves under its own viscosity until the accretion rate on the star drops below a few times $10^{-9}M_\odot$/y. At this point the photo-evaporation process becomes important, as it removes gas from $\sim 1$~AU at a comparable rate (see Alexander et al., 2014, for a review). Photo-evaporation divides the disk into an inner part, rapidly accreted onto the star, and an outer part, rapidly photo-evaporated inside-out; it explains the rapid final clearance of protoplanetary disks.}

Stone et al. (1998) pointed out that the Magneto Rotational Instability (MRI) is unlikely to act uniformly across the disk. The central part of the disk, which is opaque to radiation, should not be ionized and therefore the MRI should not operate there. This led to the concept of a ``dead zone'', a low-viscosity region within the disk. Disks with a dead zone can still be modeled with the $\alpha$ prescription, provided that $\alpha$ is function of the radial and vertical coordinates. Because in the dead zone $\alpha$ is smaller, the gas density is much larger than in the MRI active zone to ensure the smooth transport of material through the disk. In disks modeled in 1D (i.e. radial structure only) the contrast in gas surface density $\Sigma_{\rm dead}/\Sigma_{\rm MRI}$ is proportional to $\alpha_{\rm MRI}/\alpha_{\rm dead}$ (e.g. Martin et al., 2012). In 2D models ($r,z$), however, there is no simple proportionality, because the gas can flow near the surface of the disk (Bitsch et al., 2014b). 

Recent studies have shown that the MRI is likely to be suppressed almost everywhere in the disk because of non-ideal MHD effects such as ambipolar diffusion (Bai and Stone, 2013; Lesur et al., 2014; see Turner et al., 2014 for a review). Disk winds have been proposed to be the main process removing angular momentum from the disk, thus causing gas to accrete onto the central star (Bai and Stone, 2013; Bai, 2013, 2014, 2015, 2016; Gressel et al., 2015; Simon et al., 2015).  The resulting disk structure can be very different from that of an $\alpha$-disk. For instance, Suzuki et al. (2010, 2016) find that the inner part of the disk can be substantially depleted, although Bai (2016) still finds a $1/r$ surface density profile in the innermost $\sim 10$~AU, similar to that of an $\alpha$-disk. The actual surface density profile is very important for models of planetesimal formation and planet migration. For instance, a depleted inner disk would slow down (or even reverse) the radial migration of protoplanets (Ogihara et al., 2015a,b). It would also have a much less sub-Keplerian rotation than an $\alpha$-disk, favoring particle accumulation and the formation of planetesimals. 

\begin{figure}[t!]
\centering
\includegraphics[width=20pc]{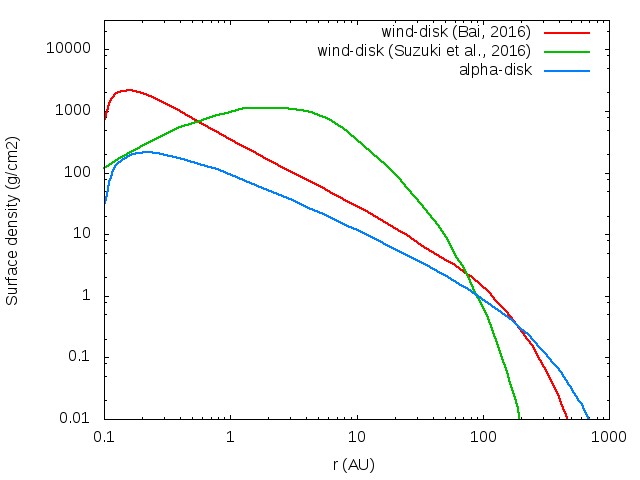}
\caption{\footnotesize Structure of different disk models, all with an accretion rate on the star of $\sim 5\times 10^{-9} M_\odot/y$. Each curve shows the surface density as a function of heliocentric distance. Two models of low-viscosity disks where transport is dominated by winds are shown, from Bai (2016) (red color; $\alpha=2\times 10^{-4}$) and Suzukui et al (2016) (green color; $\alpha=8\times 10^{-5}$). They are labeled as ``wind-disks'' to contrast with the classic $\alpha$-disk, whose density distribution is also shown in blue. In the $\alpha$-disk model the surface density of the gas scales inversely to the assumed viscosity {( here $\alpha=0.01$ is assumed)}. Note the depletion of gas in the inner part of the disk in Suzuki et al. model.}
\end{figure} 

An important question to address for a wind-dominated disk is how low the viscosity (or the turbulent strength) can be. There are sources of turbulence other than the MRI, such as the baroclynic instability (Klahr and Bodenheimer, 2003), the vertical shear instability (Nelson et al., 2013) { and the zombie-vortex instability (Marcus et al., 2015)}. Stoll and Kley (2014) estimated that the vertical shear instability can sustain turbulence with a strength equivalent to $\alpha \approx$ a few times $10^{-4}$, but more recent work (Stoll and Kley, 2016) argues that this is true only beyond $\sim$5~AU. Thus, the actual strength of the turbulence and the associated viscosity in the inner disk is still unknown. The strength of turbulence is a key parameter for planetesimal growth, because it governs the settling of particles towards the midplane. The viscosity is also the key to estimate the temperature of the disk. 

In fact, in the absence of viscous heating, the only source of heat is irradiation from the central star\footnote{{ Actually, Gressel et al. (2015) point out the existence of current layers at about $z=2H$ which should generate heat. The role of these currents on the thermal structure of the disk is still unexplored.}}. A disk that is solely heated by the star is called a passive disk (Chiang and Goldreich, 1997; Dullemond et al., 2001, 2002). A passive disk is very cold throughout its evolutionary stages, until it becomes transparent to the stellar irradiation. Chiang and Youdin (2010) estimated that the snowline (the location where the temperature is 160--170K so that water has a transition from vapor to solid state) in such a disk should be around $\sim 0.4$~AU. If this were the case, all bodies in the inner solar system should be water rich, including main belt asteroids and the Earth. While these bodies contain some water, their bulk water abundance is much less than that expected from solar abundance (about 50\% in mass; Lodders, 2003), suggesting that the snowline { was, at least during the early phases of the disk,} much father out than their location. { In fact, the mechanisms proposed to explain why the inner solar system bodies did not accrete a lot of water as the disk cooled off (Morbidelli et al., 2016; Sato et al., 2016) require that the disk was initially too hot for the existence of water-ice at 1--2 AU.} Thus, we expect that there was some non-negligible turbulent viscosity in the inner disk, even if the origin of this viscosity is still not clear. For reference, for the snowline to be at about 3~AU, as suggested by asteroid composition, the viscous transport in the disk should have been of about $3\times 10^{-8}M_\odot/y$ (Bitsch et al., 2015). If the disk had this viscous transport at some point during its evolution, its viscosity had to be large or the disk very massive. See Morbidelli et al. (2016) for a more in depth discussion. 

While it may seem tangential to planet formation, angular momentum transport and heating in protoplanetary disks is essential in determining the disk's structure.  This feeds back into particle growth and orbital migration.

\section{How did the first planetesimals form?}
\label{planetesimals}

The solid building blocks of planets are a trace component within gas-dominated protoplanetary disks. The processes that govern the growth of macroscopic particles remain poorly understood. Perhaps the most vexing problem is the growth of the first planetesimals. Once planetesimals are present in the disk, their further growth by accreting pebbles is modestly-well understood. The first population of planetesimals may serve as a planetary system's blueprint.  

In the classic model of planet accretion, dust particles settle to the midplane of the disk, collide with each other and form aggregates held together by electrostatic forces (e.g., Dominik \& Tielens 1997). Little by little these aggregates grow and get compacted by collisions, forming small planetesimals (e.g., Weidenschilling and Cuzzi 1993, Blum and Wurm 2008).

Today we know that this classic picture of planetesimal growth has severe problems (illustrated in Figure 2). When silicate grains grow to a size of about a millimeter, they start to bounce off each other instead of accreting ({ Zsom and Dullemond, 2008;} G{\"u}ttler et al., 2009). In the icy part of the disk, particles can grow up to a few decimeters in size before starting to bounce. This is called the {\it bouncing barrier}. At these sizes, particles migrate rapidly in the disk towards the star due to gas drag. This radial drift produces large relative velocities among particles of different sizes and hence disruptive collisions. Even if collisions were not disruptive and particles could continue to grow, eventually meter-size boulders would migrate so rapidly to be lost into the star before they can grow significantly further (Weidenschilling, 1977). This is the well known {\it meter-size barrier} (referred to as the ``drift barrier'' in Figure 2). 

Over the past decade new ideas have been proposed to overcome this problem. 
{ On the one hand, it has been shown (Okuzumi et al., 2012; Kataoka et al., 2013) that the regime of collisional coagulation might nevertheless work beyond the snowline because of the tendency of icy particles to form very fluffy aggregates. These aggregates, thanks to their extremely high porosity, are very resistant to collisional disruption and their radial drift in the disk is also very slow. Thus both the bouncing/breaking barrier and the drift barrier can be circumvented and km-size planetesimals can be formed after a final phase of contraction of the aggregates. This alleged formation path for planetesimals, however, should not work within the snowline, due to the reduced sticking properties of the silicate grains.

On the other hand, it has been shown that ``small'' particles (whatever their chemical nature)} tend to clump due to turbulence in the disk (Cuzzi et al., 2001; Johansen et al., 2007). Once the clump is sufficiently dense, its self-gravity can take control, holding all particles together until they settle in their common potential well and form a large planetesimal (Johansen et al., 2007; Cuzzi et al. 2010). This recipe, however, is not without problems. First, it requires turbulence within the disk. Yet the very existence of turbulence and its properties are far from being firmly established, as discussed in the previous section. Second, in order to clump under the effect of turbulence, particles must be much larger than that allowed by the bouncing barrier, at least in the inner disk. { In fact, the critical parameter governing the dynamics of a particle undergoing gas drag is its Stokes number $S_t=t_f \Omega$, where $\Omega$ is the Keplerian frequency and $1/t_f$ is the coefficient relating the acceleration of the particle due to the drag to the velocity difference between the particle and the fluid medium, i.e. $\dot{v_{\rm p}}=1/t_f (v_{\rm p}-v_{\rm gas})$.}  In the simulations by Johansen et al. (2007), particles effciently clumped have Stokes number (i.e. the product of the friction time and the orbital frequency) larger than 0.5 which, at 1 AU, corresponds to sizes of $\sim$20cm. The concentration mechanism in low-vorticity regions proposed by Cuzzi et al. (2001) seemed to operate for chondrule-sized particles, but a new reconsideration of this mechanism (Cuzzi et al., 2016), adopting an improved turbulent cascade model, also raises the particle size to 10cm or so. How particles can grow to this size in the inner disk is an open problem. 

\begin{figure}[t!]
\centering
\includegraphics[width=9.cm]{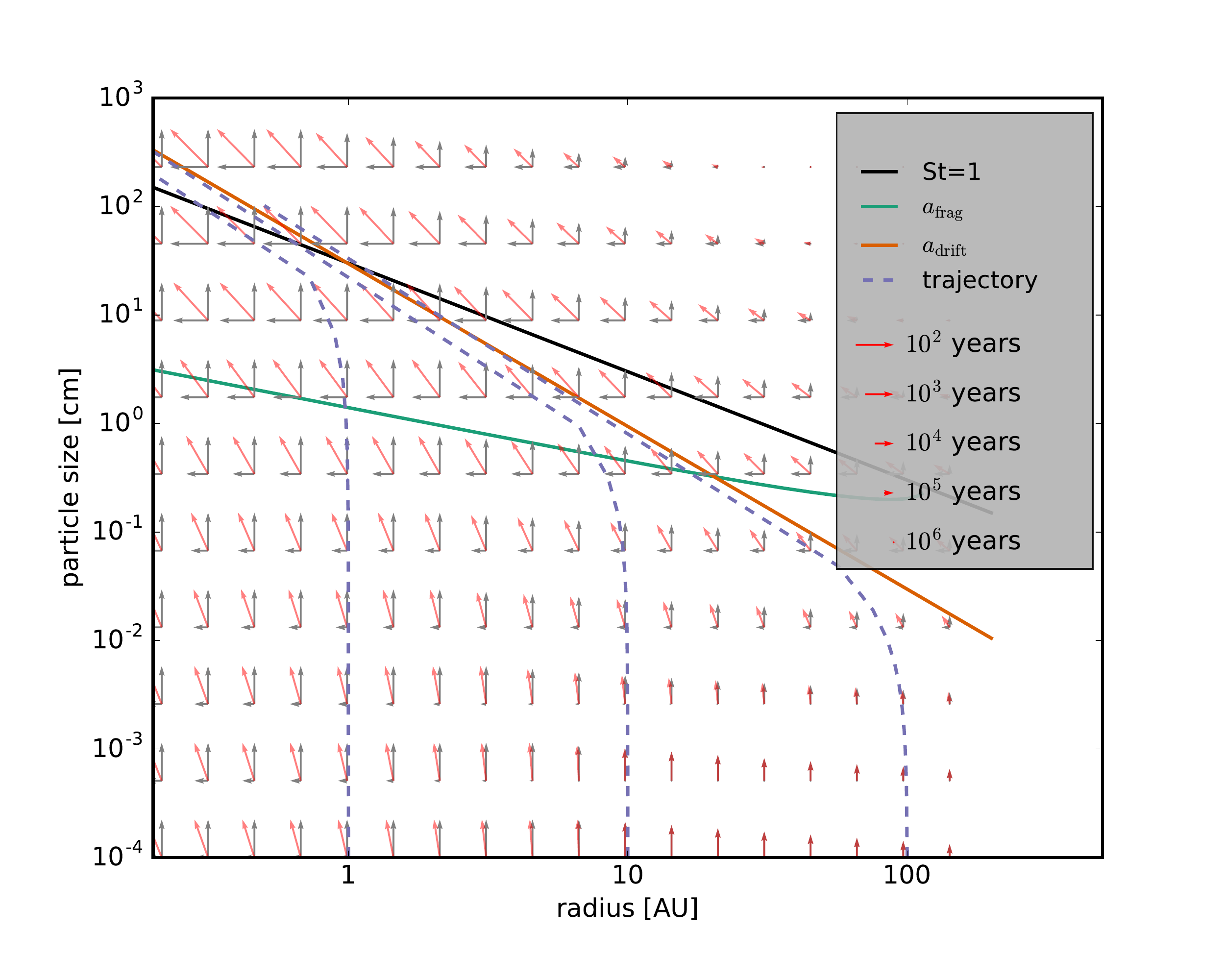}
\caption{\footnotesize Growth and drift of particles in a protoplanetary disk.  Trajectories for the growth (vertical arrows) and inward aerodynamic drift (horizontal arrows) were calculated at each point in a specific disk model, with a 1/r surface density profile and dust-to-gas ratio of 0.5\% (Birnstiel and Andrews 2014). The arrow length is inversely proportional to the relevant logarithmic timescales; i.e., to $a/\dot{a}$ (for inward drift) and $r/\dot{r}$ (growth).  The green line shows the fragmentation barrier above which growth is inhibited. The dashed/purple curves show the trajectories of particles starting at different orbital radii (calculated neglecting fragmentation). The red line is the corresponding drift barrier.  The solid black line denotes where the Stokes number is equal to 1.  From Birnstiel et al (2016). }
\end{figure}

In order to relax the hypothesis that the disk is a priori turbulent, it has been proposed that particles themselves can generate turbulence and then clump in a feedback process. Two mechanism can produce this result. The sedimentation of particles towards the midplane of the disk can drive a Kelvin-Helmoltz instability, because the particles force the midplane layer of the gas to rotate around the Sun faster than the surface layers (Weidenschilling, 1995). As shown in Johansen et al. (2006), the turbulence generated by the Kelvin-Helmholtz instability can clump the particles enough to form self-gravitating clusters. But, again, this effect depends on particle size. Small particles do not sediment very fast and therefore don't create a dense midplane layer; moreover, being more coupled with the gas, the shear effect that they generate is weaker. Effective clumping leading to local sold/gas density ratio of a few 10s is observed only for particles with Stokes number larger than 0.1, which corresponds to a size of 10cm at 1 AU. 

The second mechanism is the streaming instability (Youdin and Goodman, 2005). This is a linear instability generated by the radial drift of particles. When the local density of particles exceeds the background density (i.e. a small clump is formed due to some statistical density fluctuation), the collective drag force exerted by the clump on the gas makes the disk rotate faster; as a consequence, the clump as a whole drifts inwards slightly more slowly than individual particles. This creates a sort of traffic jam. Other particles drifting inward at the nominal speed catch up to the clump, enhance the local overdensity of solids, and further decelerate the radial drift of the clump. Thus, the local accumulation of particles speeds up, and so forth in a positive feedback. Yet the streaming instability is not without problems. If particles are relatively large (Stokes number larger than 0.1), clumping is effective only if the solid/gas surface density ratio is larger than 3\% (so, enhanced compared to solar value of 1\%; Johansen et al., 2009). For chondrule-size particles in the inner disk, the solid/gas ratio has to be increased above 8\% (Carrera et al., 2015). 

Models of particle coagulation and radial drift (Lambrechts and Johansen, 2014) show that both the size of available particles and the solid/gas ratio decrease in time even as the gas surface density decays. It is unclear whether the conditions for the streaming instability can ever be reached. According to Drazkowska and Dullemond (2014) the streaming instability may only be operational in the outer disk for two reasons: (i) beyond the snowline, the sticky properties of ice allow the formation of larger particles and, (ii) for a given particle size, the Stokes number is larger than in the inner disk. { In the inner disk the streaming instability may only operate in a relatively narrow ring (Drazkowska et al., 2016) where pebble pile-up if the radial surface density profile of the disk is sufficiently shallow.}  

The high solid/gas ratio required for chondrule-size particles to clump in the asteroid belt can only be reached in the late stage of the disk, after a substantial fraction of the gas has been removed (Throop and Bally, 2005) and {\it if} particles are regenerated by some process, for instance in planetesimal-planetesimal collisions. 

The analysis of meteorites shows indeed that there were at least two generations of planetesimals in the inner disk. The parent bodies of iron meteorites formed early, in the first My (Kruijer et al., 2012) { after the formation of the first solids (the Calcium-Alluminium Inclusions -CAIs- dated to have formed 4.568 Gy ago; Bouvier et al. 2007a; Burkhardt et al. 2008)}. In contrast, the parent bodies of chondritic meteorites formed after 3-4 My (Villeneuve et al., 2009). It is unclear whether there was a gap in time in planetesimal formation and even whether the first and last generations of planetesimals formed in the same place. Today all meteorites come from the asteroid belt, but the asteroids may have been trapped into the belt from different original reservoirs (Bottke et al., 2006; Levison et al 2009; Walsh et al., 2011; Raymond et al 2016). Some authors have proposed that chondrules are the outcome of collisions among massive objects (Libourel et al., 2006; Asphaug et al., 2011; Johnson et al., 2015). All this may be suggestive that the first planetesimals formed only at select locations (near the disk's inner edge or beyond the snowline) and that elsewhere planetesimals only formed later, as the protoplanetary disk evolved towards the debris disk phase. But this is still far from a consolidated view. It is not clear, for instance, why the second generation asteroids would have only accreted chondrules and not a spectrum of { collisional debris with different physical and chemical properties}. 

Once planetesimals are formed, as long as they are embedded in a disk of gas and drifting particles { and they are big enough (typically larger than 100km)}, they can keep growing by accreting the particles via a combination of gravitational deflection and gas drag (Ormel and Klahr, 2010; Johansen and Lacerda, 2010; Lambrechts and Johansen, 2012, 2014; { Guillot et al., 2014}; Johansen et al., 2015; Levison et al., 2015; Visser and Ormel, 2016). This process is called {\it pebble accretion} and it is now fairly well understood and recognized to play a fundamental role in planet formation. Unfortunately, { by definition,} pebble accretion operates only after that the first planetesimals are formed. Thus, our poor understanding of the primary planetesimal formation process, discussed above, blocks our understanding of when, where and in which numbers, protoplanets formed in the disk.

\section{How fast, and in what direction, do planets migrate?}
\label{migration}

Growing planets interact gravitationally with the protoplanetary disk. They create a spiral density wave in the gas distribution that is stationary in the reference frame rotating with the planet (Ogilvie and Lubow, 2002). In turn, this non-axisymmetric density distribution exerts a torque on the planet. As a result, the angular momentum of the planet's orbit changes and the orbit expands or contracts depending on the sign of the torque. In other words, the planet undergoes a radial migration. It can be demonstrated that in disks whose surface density is a power law of $1/r$ the torque is negative and hence the planet migration is directed towards the central star (Ward, 1997). For low-mass planets, the migration speed scales linearly with the planet's mass (Goldreich and Tremaine, 1980; Lin and Papaloizou, 1986). Thus, { planet migration can be neglected up to, at most, Mars-mass embryos, such as the precursors of terrestrial planets (Wetherill, 1990). According to Tanaka et al. (2002), the migration timescale of a Mars-mass body at 1 AU in a MMSN disk is 1~My. Given that these bodies took themselves a few My to form (Dauphas and Pourmand, 2011) it is likely that they emerged in a disk less massive than the MMSN, and therefore did not migrate significantly during the disk phase. For this reason, terrestrial planet formation models typically neglect radial migration. } But for proto-planets of one to few Earth masses, such as the growing cores of giant planets, migration cannot be ignored. 

The discovery of hot Jupiters (Mayor and Queloz, 1995) was welcomed as a proof of planet migration (Lin et al 1996). But it was soon understood that the hot Jupiters are the exception rather than the rule, as most extrasolar giant planets have orbital radii of 1-2 AU (Butler et al 2006; Udry \&  Santos 2007). In order to reproduce the observed distribution of planets, planet synthesis models (e.g. Ida and Lin, 2004a,b; Alibert et al., 2005) had to  artificially reduce the speed of migration by one to two orders of magnitude relative to theoretical expectations.

No mechanism to globally reduce planets' migration speed across the disk has been found. Detailed studies have identified effects that can reduce, stop and even reverse migration, but only at specific locations. Masset et al. (2006) found that a positive surface density gradient in the disk can act as a trap for migrating planets. However, the only place where such a gradient is expected is at the inner edge of the disk (or at the boundary between the MRI active region and the dead zone of the disk, which is also very close to the star).  The mechanism of Masset et al (2006) can explain why planets don't fall on the star but accumulate on small semi-major axis orbits; however, it cannot explain how the planets that are a few AU away could have avoided migrating closer to the star. 

\begin{figure}[t!]
\centering
\includegraphics[width=9.cm]{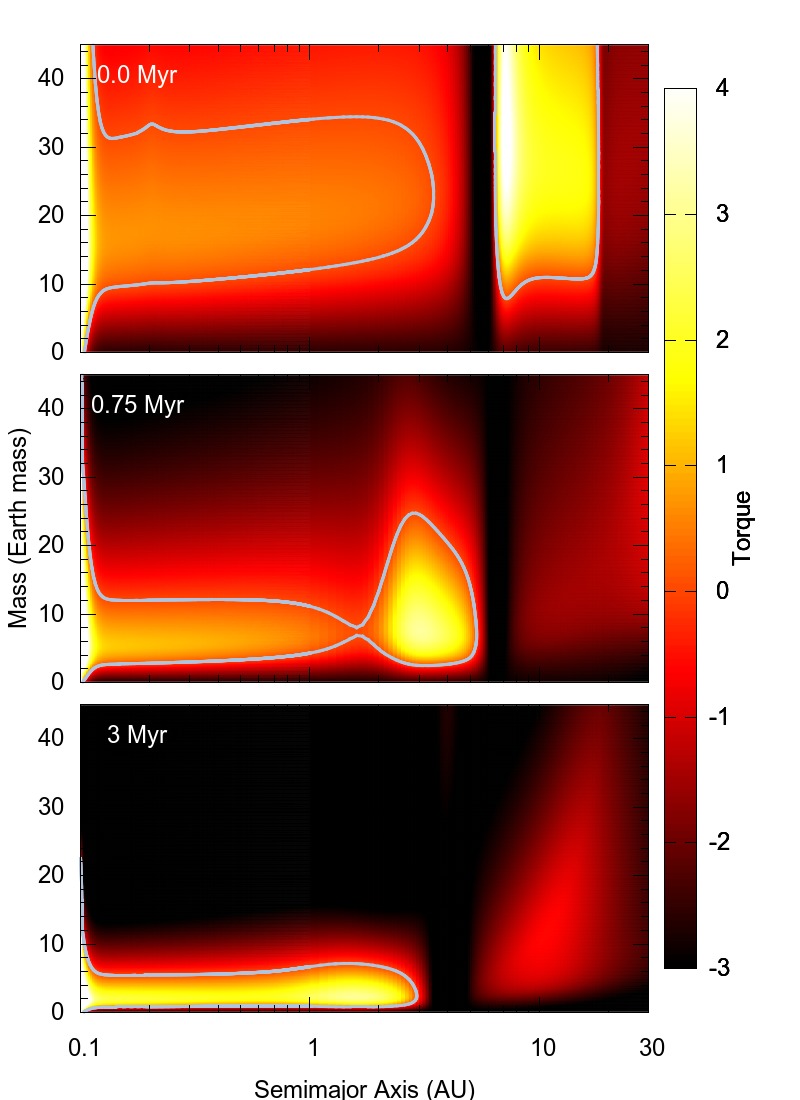}
\caption{\footnotesize A migration ``map'' for an evolving protoplanetary disk.  In each snapshot, the color denotes the torque felt by a planet of a given mass at a given orbital radius.  A positive torque will cause the planet to migrate outward and a negative torque to migrate inward.  Contours follow zero-torque locations and enclose regions of outward migration.  In the first snapshot there are two distinct zones of outward migration, which merge into one that moves inward and to lower masses as the disk evolves.  The disk was modeled using the $\alpha$ prescription (Shakura \& Sunyaev 1973) with $\alpha=5\times 10^{-3}$.  From the disk model of Bitsch et al (2015).  }
\label{Bitsch}
\end{figure}

Paardekooper and Mellema (2006) found a positive torque that arises in non-isothermal disks with steep temperature gradients. This torque can exceed the negative one exerted by the spiral density wave mentioned above, thus forcing the outer migration of the planet. Several years of studies led to a quantitative analysis of this torque (Paardekooper et al., { 2010}, 2011). When applying these torque formulae to the structure of an $\alpha$-disk ({ Hasegawa and Pudritz, 2011; Hellary and Nelson, 2012;} Bitsch et al., 2015), it is found that the outward migration region is confined in radial range and planet mass (see Fig.~\ref{Bitsch}). In an massive young, warm disk, the main region of outward migration extends from 3 to 9 AU and encompasses planets between 5 and 50 Earth masses (see Figure 3). This is good news, because it means that { a planet} in this mass range, migrating from the outer disk, would stop at the outer border of this region. But later on, when the disk becomes cooler and less massive, the region of outward migration shifts inwards and to smaller planet masses (Fig.~\ref{Bitsch}). Bodies more massive than 10 Earth masses are released to rapidly migrate inward. For low-viscosity, MRI inactive disks (see Section~\ref{disk}), the maximal planet mass of the outward migration region is even smaller. { For multiple planets the situation may be even worse. Mutual interactions among the planets may raise their orbital eccentricities and in this case the aforementioned positive torque is strongly reduced, promoting inward migration (Bitsch and Kley, 2010; Hellary and Nelson, 2012).} 

The currently-favored paradigm of giant planet formation proposes that 10-20 Earth-mass cores serve as the seeds for gas giants (Pollack et al 1996). There is a delay between the core's growth and the onset of rapid gas accretion. During this interval massive cores may migrate inward and, following current models, will reach the inner edge of the disk before becoming giants (Coleman and Nelson, 2014). With our current understanding of migration, for giant planets to form and still be at a few AU from the star when the disk disappears, Bitsch et al (2015b)  found that their cores must start their growth at 15--20~AU and relatively late during the disk lifetime. Yet issues remain regarding this result.  Because the outer disk is flared, both the streaming instability and pebble accretion are more efficient closer in than farther out. Thus, it is difficult to understand why giant planet cores could not form between the original snowline location (say 3-5 AU) and 15 AU, if they could form beyond this threshold distance. One possibility is that planets that formed closer-in or earlier were engulfed into the central star and that the planets that we see today are the ``last of the Mohicans'', namely those that formed the last and the farthest, so that they could not reach the star before disk removal (Lin, 1997; Laughlin and Adams, 1997). But this is difficult to believe: the abundance of numerous close-in super-Earths argues that planets stop migrating at the inner edge of the disk (Masset et al., 2006) and don't fall onto the star; indeed, disks have inner edges that stop inward migration (Romanova et al 2003; Bouvier et al 2007b). Moreover, in our Solar System it is hard to conceive that numerous planets migrated through the inner disk without leaving a trace of their passage. The case of our Solar System may be helped by the fact that the contemporary presence of Jupiter and Saturn in nearby orbits reverses the migration direction of these giant planets (Masset and Snellgrove 2001). But even in this case Bitsch et al. (2015b) could not achieve the required Jupiter-Saturn configuration if these planets started growing within 23 AU. 

It is likely that we are still missing a fundamental piece of the migration story.  One idea is that, if disks are much less viscous than expected (see Sect.~\ref{disk}) relatively small planets could open gaps in the gas distribution. They would then migrate in a different regime, known as the Type-II regime (Lin and Papaloizou, 1986), in which the migration speed is equal to the radial speed of the gas due to its viscous evolution. If the viscosity of the disk is small, this speed is also small. However, this idea seems naive. First, to open a gap in the gas distribution, it is not enough that the torque exerted by the planet overcomes the viscous torque within the disk; it is also needed that the planet has a Hill radius larger than the scale height of the disk (Crida et al., 2006; Morbidelli et al., 2014), otherwise the gas flows into the gap from the vertical direction. Second, Type-II migration is an ideal process, but in reality gap-forming planets do not migrate at the viscous radial-drift speed ({ Duffell et al., 2014}; Durmann and Kley, 2015). In particular, in the low-viscosity regime migration is faster than the radial drift of the gas. 

Additional positive torques exerted on the planets have been searched for in the last years. Benitez-Llambay et al. (2015) found that a few Earth-mass planet with a very high temperature because of a vigorous accretion of solids can migrate outwards. However, this effect wanes in the mass regime of giant planet cores, so it does not help the giant planet formation problem (Bitsch et al., 2015b). Fung et al. (2015) claimed to have found a new positive torque arising only in 3D simulations of planet-disk interactions. However, this torque seems to disappear when simulations are conducted with a sufficiently high resolution (Masset, private communication). Thus, nothing promising is in view at the current time. 

Maybe it is the disk structure that is misleading us (Coleman and Nelson, 2016). As we saw in Sect.~\ref{disk} the accretion $\alpha$-disks that are usually considered may not be realistic. If the new disk-wind paradigm of transport creates a global positive surface density gradient in the inner few AUs of the disk, as advocated by Suzuki et al. (2010), the migration of planets can be substantially modified  (Ogihara et al., 2015a,b). However, it is not yet sure that this disk structure is real (see Bai, 2016). More work on the disk structure is needed.

\section{What is the origin of the Solar System's peculiar orbital structure?}
\label{SSstructure}

The Solar System is characterized by a tri-modal structure. In its inner part there are the terrestrial planets, rocky and relatively small. Farther out, there is the asteroid belt, which carries cumulatively a negligible mass (about $5\times 10^{-4}$ Earth masses). Beyond 5 AU lies the realm of the giant planets. The formation of the terrestrial planets requires that, during the lifetime of the protoplanetary disk, a set of planetary embryos formed in that region, with approximately the mass of Mars. Mars itself could be a surviving embryo (Dauphas and Pourmand, 2011). The small mass ratio between Mars and the Earth suggests a substantial depletion in the total mass of solid beyond 1~AU and across the asteroid belt (Hansen, 2009; Raymond et al 2009), established very early during the history of the Solar System. In contrast, the formation of the giant planets requires the accretion of solid cores of 10--20 Earth masses before the removal of gas. 

How this tri-modal structure was set is still an issue of open debate. Walsh et al. (2011) proposed that the depletion in solid mass beyond 1 AU was caused by a specific pattern of Jupiter's migration. Keeping in mind the caveats discussed in the previous section, it is legitimate to postulate that Jupiter initially migrated inward, when it was basically alone in the disk, and then reversed migration direction when Saturn formed (Masset and Snellgrove, 2001; Morbidelli and Crida, 2007; Pierens and Nelson, 2008; Pierens and Raymond, 2011; discussed at length in Raymond \& Morbidelli 2014). If this inward-then-outward migration history brought Jupiter down to $\sim 1.5$~AU before bringing it into the outer disk again, the region beyond 1~AU would have been strongly depleted in mass (Fig. 4). Indeed, the simulations of Walsh et al. show that the terrestrial planets would have formed with a mass distribution very similar to the one observed (see also Jacobson and Morbidelli, 2014), and the asteroid belt would have acquired a structure consistent with the current one (see Deienno et al., 2016). If this scenario, dubbed ``Grand Tack'' is correct, then the Solar System had initially just a bi-modal structure, with Mars-mass planetary embryos inside of the snowline and giant planet cores outside; the tri-modal structure (i.e. the depletion between 1 AU and the orbit of Jupiter) was generated by Jupiter's migration (Fig. 4).

However, even this bimodal structure is puzzling. Assuming that planetesimals formed everywhere in the disk with comparable masses (but see Sect.~\ref{planetesimals}), the subsequent process of planet growth by pebble-accretion should favor the bodies closer to the Sun (Ida et al., 2016). In other words, giant planet cores should have formed in the inner disk and Mars-mass embryos in the outer disk! The only exception to this is if the pebble sizes drastically changed at the snowline. Morbidelli et al. (2015) postulated that beyond the snowline the pebbles were dm-size objects made of ice and silicate grains. When these pebbles drifted across the snowline the ice sublimated: 50\% of the solid mass was lost and small (mm-size) silicate grains were released { (See also Saito and Sirono, 2011)}. If there was enough turbulence in the disk to make the particle layer thicker than the Hill radii of the planetesimals (a weak turbulence with $\alpha=10^{-4}$ would have been sufficient) the accretion of small grains would have been much less efficient than the accretion of large pebbles. Hence, two initially equal-mass planetesimals on either side of the snowline would have grown at drastically different rates: by the time the outer one reached 20 Earth masses, the inner one would have barely reached Mars-mass (Morbidelli et al., 2015; Ida et al., 2016). This explanation of the bimodal structure of the Solar System, however, is not unique: if planetesimals inside the snowline were smaller for some reason (see Sect.~\ref{planetesimals}) than planetesimals beyond the snowline, the latter would have grown much faster by pebble-accretion even without a drastic change in pebble size across the snowline. 

But let's take a step back and wonder whether the tri-modal structure of the Solar System could have been primordial (i.e. not the consequence of a Grand Tack-like migration of Jupiter). Starting from a smooth population of planetesimals Levison et al. (2015b) obtained a concentration of mass inside of 1 AU through pebble-accretion. Their model involves a number of assumptions, three of which appear particularly critical to us. First, it is assumed that there is no mass flux in solids coming from beyond the snowline; second, pebbles form by condensation and coagulation throughout the inner disk; third, the disk is flared everywhere. The first and the second assumptions together have the consequence that, the closer a planetesimal is to the snowline (but inwards of its location), the smaller the pebble-flux that crosses its orbit. The third assumption makes the pebble-accretion process more efficient with decreasing heliocentric distance. 

These three assumptions may not be well justified. Preventing a flux of pebbles from the outer disk requires the formation of a strong positive gas surface density gradient (i.e. a pressure ``bump'') at the snowline. It has been proposed { in the context on an analytic 1D (radial) disk model} (Kretke and Lin, 2007) that the snowline can create a pressure bump because the condensation of icy grains could strongly reduce the ionization of the gas, quench its turbulent viscosity, and increase the gas surface density beyond the snowline. However, { 3D numerical simulations of disk structure in the presence of radial and vertical viscosity transitions} (Bitsch et al., 2014b) show that this is true only if ${{\rm d}\log(\nu)\over{{\rm d}\log(r)}}< -28$, where $\nu$ is the viscosity. It is very unlikely that the condensation of icy grains could have such a dramatic effect on the viscosity. As for the second assumption, it has been shown that the direct condensation of gas is not relevant after about half a viscous timescale (Morbidelli et al., 2016). In fact, after this time, the gas drifts inward faster than the radial displacement of the condensation lines. So when the temperature drops at a given location the local gas should not carry any condensable species because that species would have already condensed farther out. Concerning the third assumption, a warm disk with a snowline at $\sim 5$~AU, as in the Levison et al. model, must be dominated by viscous heating in its inner parts (see Sect.~\ref{disk}), but viscously dominated disks are not flared ({ Kenyon and Hartmann, 1987; Chiang and Goldreich, 1997;} Bitsch et al., 2014). Having a flared disk in the inner part would requires an ad-hoc radial-dependence of the dust/gas ratio (Levison et al., 2015b). 

\begin{figure}[t!]
\centering
\includegraphics[width=9.cm]{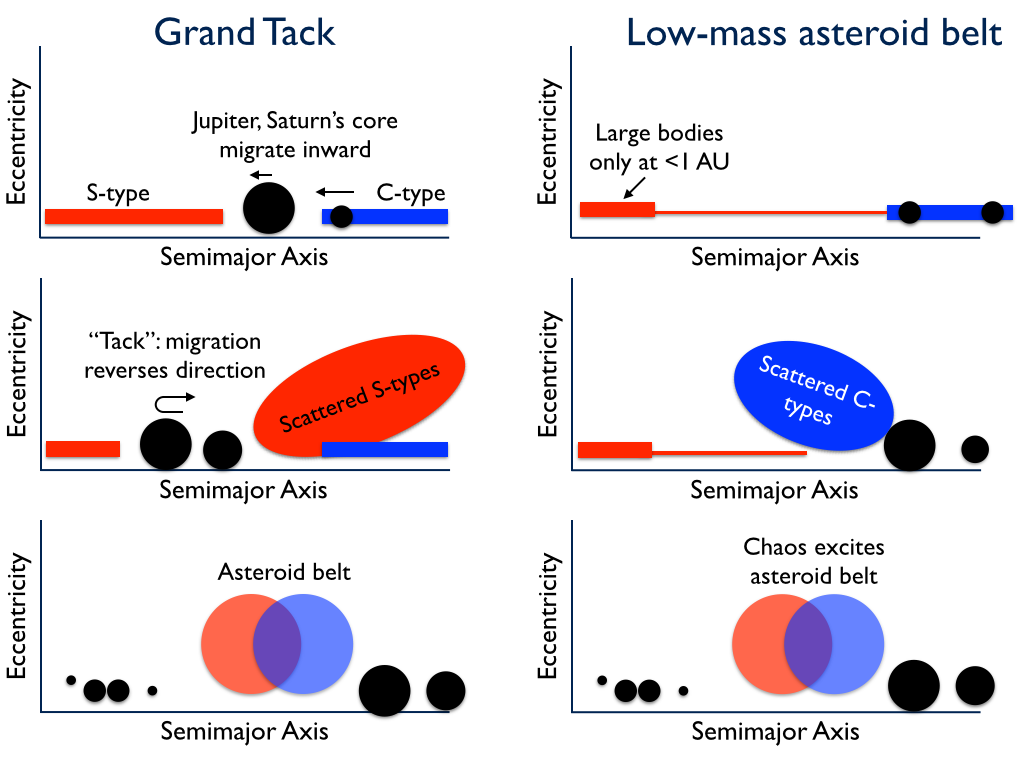}
\caption{\footnotesize Cartoon representation comparing the Grand Tack model (left) with a model in which the asteroid belt was low-mass even at early times (right).  While the evolution of the two models is quite different, each can match the present-day inner Solar System.}
\label{GT}
\end{figure}

Nevertheless, the possibility that the solid mass was concentrated within 1 AU should be taken seriously. As discussed in Sect.~\ref{planetesimals} it is possible that the streaming instability formed a first generation of planetesimals in just two places: near the inner edge of the disk (although it is unclear if this could extend out to 1 AU) and beyond the snowline. If this was the case, obviously pebble accretion could only operate in these regions. Asteroids may have only formed later, as a low mass second-generation population, from the collisional debris of the first generation. Although this view is far from being established, it would explain the tri-modal structure of the Solar System without the need for the Grand Tack. 

Not being able to decide a priori which view is the most realistic for the formation of the Solar System, we can turn to observational constraints to discriminate them. Izidoro et al. (2015c) pointed out that, in the absence of Jupiter's migration, an initially low-mass asteroid belt should have remained dynamically cold, i.e. most asteroids should have today low eccentricity, low inclination orbits. This contrasts with the observed distribution. However, in a more recent work (Izidoro et al., 2016) it was found that, if the giant planets had a secular chaotic motion (which might have happened, for instance, if Jupiter and Saturn spent some time in their mutual 1/2 resonance after the removal of the gas from the protoplanetary disk), the orbital distribution in the asteroid belt could have been excited by the stochastic stirring exerted by a multitude of secular resonances. Thus, the asteroid belt does not seem to be diagnostic of what really happened, given that its structure can be explained by both the Grand Tack model and by chaotic stirring (Fig. 4). 

Therefore, we probably need to turn to the terrestrial planets for more insight. The amount of water-rich material incorporated by the terrestrial planets in the two scenarios may be significantly different, although this has not been thoroughly explored. Another issue to investigate is the final orbital excitation of the terrestrial planets. Levison et al. (2015b) investigated the growth of terrestrial planets from a local annulus of planetary embryos and planetesimals. They find that the final terrestrial planet systems have angular momentum deficits (AMD: a mass-weighted mean of the eccentricities and inclinations of the planets; Laskar, 1997) that is 2 to 10 times that of the real terrestrial planets. The reason for this is that the systems of planetary embryos deplete nearly all planetesimals before having any major orbital instability that allows embryo-embryo merging. This leaves behind insufficient mass in planetesimals to significantly damp the orbits of the final planets (O'Brien et al 2006; Raymond et al 2006). If this is confirmed by future studies including also the ejection of debris in embryo-embryo collisions, it could become a strong argument in favor of the Grand Tack hypothesis (or another mechanism that injects planetesimals into the inner disk). In fact, during the inward migration of Jupiter in the Grand Tack model, a large number of planetesimals are transported from the asteroid belt into the terrestrial planet region (Walsh et al., 2011). This repopulates the terrestrial planet region of planetesimals, which then damp the eccentricities and inclinations of the forming planets by dynamical friction, allowing the final system to meet the AMD constraint (Jacobson and Morbidelli, 2014).

\section{Are super-Earths and terrestrial planets the same or different?}
\label{superearthssect}

So-called ``Super-Earths'' are the most abundant class of planets known to date.  Exoplanet searches have found that roughly half of all Sun-like stars have one or more planets with radii of 1-4 Earth radii and orbital periods shorter than $\sim$100 days (Mayor et al 2011; Howard et al 2012; Fressin et al 2013; Petigura et al 2013).  

What can we learn about planet formation from the population of super-Earths?  (For simplicity, in this discussion we refer to all exoplanets smaller than $4 R_\oplus$ and with orbital periods shorter than 100 days as {\it super-Earths}.) Super-Earths are often found in multiple systems, usually in compact orbital configurations with low eccentricities and low mutual inclinations (Lissauer et al 2011, Fabrycky et al 2014; Tremaine \& Dong 2012; Fang \& Margot 2012).  Their orbital separations -- measured in orbital period ratio or in mutual Hill radii -- are similar to those among the Solar System's terrestrial planets (Fang \& Margot 2013).  It is thus tempting to imagine that these planets formed by the same processes as the terrestrial planets, via in-situ growth of rocky planetesimals or planetary embryos.  Invoking the ad-hoc existence of a large population of embryos close to their stars can indeed reproduce some properties of the observed super-Earth systems (Hansen \& Murray 2013). See Fig. 5.  However, the in-situ accretion hypothesis fails for several reasons.  First, if super-Earths formed locally, then one can construct a ``minimum-mass extra-solar nebula'' (MMEN) to understand the distribution of solids close to the star (Raymond et al 2008; Chiang \& Laughlin 2013; note that Kuchner, 2004 built an analogous nebula using close-in giant exoplanets). But the MMEN is both extremely dense (1-2 orders of magnitude more massive than the nebula inferred from the Solar System or from observations of disks around young stars; Andrews et al 2009; Williams \& Cieza 2011) and inconsistent among different systems, with a broad range of surface density profiles (Raymond \& Cossou 2014; Schlichting, 2015).  It is difficult to reconcile such a nebula with the structure of an actual protoplanetary disk.  Second, whereas migration of planetary embryos { had a limited effect} for Mars-mass planetary embryos in the primordial Solar System, migration cannot be ignored in the case of super-Earths. Given the super-Earths' higher masses and closer proximity to their stars (which likely translates into higher gas densities), the timescales for orbital migration are extremely short (Ogihara et al 2015a).  In fact, in the in-situ accretion model the inner disk was so dense that even the timescale for in-spiraling of embryos due to aerodynamic drag was shorter than the disk lifetime (Inamdar \& Schlichting 2015).  This means that, even if the super-Earths formed in-situ, their orbits must have changed. In other words, super-Earths could not have formed in-situ in the strictest sense.

It is nevertheless possible that super-Earths grew from solids that were transported inward. Yet the size scale of inward-transported solids remains debated. { Moriarty and Fischer (2015) proposed that super-Earths formed from in-situ planetesimals which efficiently accreted pebbles flowing from the outer disk.} The model of Chatterjee \& Tan (2014; see also Chatterjee \& Tan 2015 and Hu et al 2016) { instead does not require the pre-existence of planetesimals in the inner disk, but} proposes that pebbles drifted inward and were trapped at the inner boundary of the dead zone (Fig. 5). When enough mass was trapped, { gravitational} instability was triggered that led to the formation of a planet.  The dead zone receded, pebbles were trapped farther out at this new location, and the process continued.  The pebble drift model is appealing in its simplicity yet has only been explored in simple analytical terms.  While promising, the actual trapping, growth, and subsequent orbital migration of super-Earths by this mechanism needs further study.

An alternate model proposes that the inward transport occurred at larger size scales via proto-planet migration (Terquem \& Papaloizou 2007; { Ogihara and Ida, 2009;} McNeil \& Nelson 2010; Ida \& Lin 2010; Cossou et al 2014). See Fig. 5. Planetary embryos similar to those that built the cores of the giant planets formed at orbital radii of a few to a few tens of AU.  Embryos migrated inward until they reached the inner edge of the disk.  Embryos piled up and their orbits formed a chain of mean motion resonances (sometimes with a planet pushed interior to the disk's inner edge).  When the gas disk dissipated (and its dissipative effects disappeared), { some of the resonant chain destabilized, particularly those where the number of planets of the resonant chain was larger than a threshold value (Matsumoto et al. 2012). These destabilized systems suffered a late stage phase } of giant collisions between embryos.  With very simple assumptions the surviving planets quantitatively match the observed distributions (Cossou et al., 2014; Ogihara et al. 2015a). This also predicts that some systems of super-Earths -- those that avoided late instabilities -- should survive in resonant chains such as the recently-identified 4-planet resonant chain in the Kepler-223 system (Mills et al. 2016).

\begin{figure}[t!]
\centering
\includegraphics[width=9.cm]{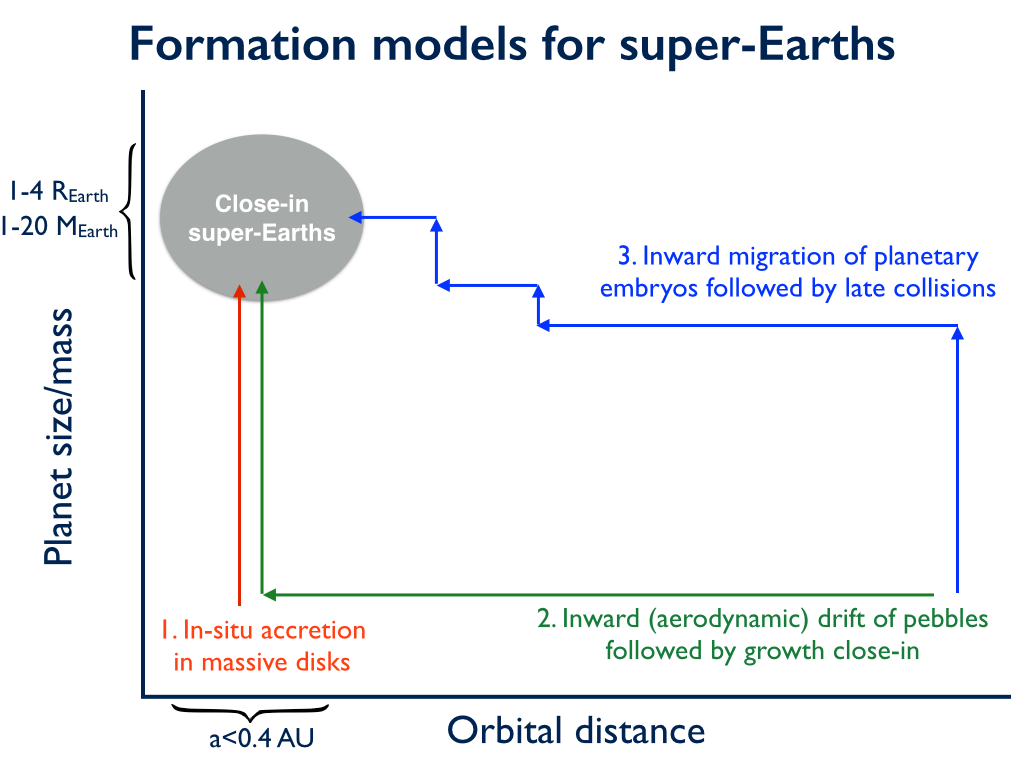}
\caption{\footnotesize Illustration of the different formation models for super-Earths. }
\label{superearths}
\end{figure}

While effective and simple, the migration model is built on two other models that have their own uncertainties.  First, it relies on the rapid formation of a population of planetary embryos (see Lambrechts \& Johansen 2014; Morbidelli et al 2015). The formation of super-Earths in the context of realistic pebble accretion of embryos has not been simulated self-consistently to date. Second, the migration model is built on the current generation of migration maps (e.g., Bitsch et al 2015), which are themselves at the mercy of the structure and evolution of the disk (see sections 2 and 3).  This illustrates the interdependence of the different problems in planet formation.

We now turn our attention to the question of similarities and differences between super-Earths and the Solar System's terrestrial planets. We have access to a plethora of data regarding the terrestrial planets including exquisite orbital constraints on the inner and outer planetary system, geochemical measurements (for Earth and Mars), seismological data (for Earth), and a host of other information.  While we can imagine extrapolating our understanding of our Solar System, we { cannot hope to obtain similar data for planets around other stars in a forseable future}.

The key question remains: in making an analogy between our terrestrial planets and super-Earths, would we be comparing apples with apples?

Regardless of how they form, super-Earths grow much faster than Earth. Earth's accretion concluded (i.e., it reached $\sim$99.5\% of its final mass) with a final giant impact that is thought to have created the Moon (Benz et al. 1986; Canup \& Asphaug 2001) roughly 50-150 Myr after the start of planet formation (Kleine et al 2009; Jacobson et al 2014).  Accretion times close to the star are extremely short, and simulations show that Earth-sized or larger planets grow within 0.1-1 Myr (Lissauer 2007; Raymond et al 2007; Bolmont et al 2014; Ogihara et al 2015a).  And any migration must of course have taken place during the gaseous disk phase, which lasted a few Myr (Haisch et al 2001; Hillenbrand 2008; but see Pfalzner et al 2014 for a different view). The orbital spacing of super-Earths is suggestive of a final phase of instability (Cossou et al 2014; Pu \& Wu 2015), which may have been triggered by the dissipation of the gaseous disk ({ Ogihara and Ida, 2009}) and would have concluded within a few million years. 

Raymond et al (2014) proposed a distinction between super-Earths and ``giant embryos'', the difference being that giant embryos would have been fully-formed while the gas was still present in the disk.  Perhaps a clearer division would be between planets that become massive enough to migrate while the disk is still present and those that do not.  The vast majority of observed super-Earths are massive enough that the inferred migration timescales are much shorter than the typical disk lifetime (Rogers et al 2011; Swift et al 2013).  However, a handful of systems host planets so small that their migration timescales are very long (e.g., the Kepler-444 system; Campante et al 2015).  In principle, very small planets must have been built from solids that did not migrate; rather, they  either accreted in-situ or drifted inward as pebbles.

The Solar System's terrestrial planets were built from embryos { that presumably did not migrate significantly}, whereas super-Earths grew from migrating embryos.  One might imagine that variations in the protoplanetary disk mass could put super-Earths and the terrestrial planets in the same context. Low-mass disks should form small embryos that do not migrate { much} whereas high-mass disks should form larger embryos that migrate { significantly} (e.g., Kokubo \& Ida 2002).  Perhaps the Solar System simply formed from a lower-mass disk than most of the super-Earth systems.

We can qualitatively test this idea using planets observed around different types of stars, because the disk mass is thought to scale (to some degree) with the stellar mass (e.g., Williams \& Cieza 2011; see discussion in Raymond et al 2007).  Indeed, Mulders et al (2015) found that the population of super-Earths orbiting M stars is statistically different than that orbiting FGK stars.  As expected from our simple arguments, M stars host more small super-Earths are fewer large ones compared with FGK stars (with the division between ``small'' and ''large'' set at $2.8 R_\oplus$).  However, the M stars' super-Earths are still easily massive enough to have migrated. 

If even M stars' low-mass disks produce super-Earths, then some other factor is needed to explain the inner Solar System's very small embryos. Jupiter may be the key. Terrestrial embryos grew from pebbles that drifted inward through the disk, and the source of pebbles moved outward in time (Lambrechts \& Johansen 2014; Morbidelli et al 2015).  When Jupiter's core reached $\sim 20 M_\oplus$ it became a barrier to the inward flux of pebbles.  This starved the terrestrial embryos, leaving them too small to migrate. If the disk had been more massive then the terrestrial embryos could have reached the migration mass faster. And in a low-mass disk without a Jupiter, the inward-sweeping of the snowline could have delivered more material to the terrestrial embryos as the disk evolved (Morbidelli et al 2016).  

By these arguments, a system would need two conditions to be met to avoid forming migrating embryos.  First, the disk mass must be below a certain threshold to avoid forming migrating terrestrial embryos very quickly.  Second, a gas giant must form exterior to the terrestrial zone early enough to block the pebble flux and starve the inner embryos.  Of course, these conditions are linked because giant planet cores are also migrating embryos, and cores that end up at $\sim$5 AU may plausibly originate across the entire disk (Bitsch et al 2015b; Raymond et al 2016).  

Observations remain difficult to interpret. By combining radial velocity and transit data, bulk density measurements have been made for roughly 100 super-Earths (see Marcy et al 2014).  Most planets larger than 1.5-2 $R_\oplus$ have low measured bulk densities and may thus be ``mini-Neptunes'' rather than ``super-Earths'' (Weiss \& Marcy 2014; Lopez \& Fortney 2014; Rogers 2015; Wolfgang \& Lopez 2015).  Most planets smaller than $\sim 1.5 R_\oplus$ have bulk densities expected of rocky planets (Dressing et al 2015; but exceptions exist -- e.g., Jontof-Hutter et al 2014).  Of course, given the significant error bars and the uncertainty in planetary building materials, the detailed compositions of even the best-measured super-Earths remain poorly-constrained (e.g., Adams et al 2008).  

The meaning of the transition between super-Earths and mini-Neptunes at $R \approx 1.5 R_\oplus$ ($M \approx 5-6 M_\oplus$) remains unclear.  This is unlikely to mark the transition between migrating and non-migrating embryos because there is no clear correlation between orbital radius and planet size. Perhaps $6 M_\oplus$ is simply the critical mass above which cores can accrete gas from the disk (see, e.g., Ikoma et al 2001; Hubickyj et al 2005). It is interesting to note that Izidoro et al (2015b)'s simulations to reproduce the ice giants were by far the most successful if the icy building blocks were $\sim 6 M_\oplus$ in mass.  Or perhaps this transition is related to the mass above which atmospheres are more easily retained during the late stage of giant impacts (e.g., Lee et al 2014; Inamdar \& Schlichting 2015, 2016).  This remains an active area of research.

We still have not addressed the question of why there are no super-Earths in the Solar System. Indeed, the absence of terrestrial planets interior to Mercury remains a mystery in itself. This absence -- as well as Mars' small mass -- can be explained if the initial distribution of embryos in the Solar System only extended from roughly 0.7 to 1 AU (Hansen, 2009).  While significant work has been devoted to understanding the origin of the mass depletion beyond 1 AU, relatively few studies have addressed the very inner Solar System.

The absence of super-Earths can be explained relatively naturally by the migration model.  Imagine that the first large planetary embryo that formed migrated to a zero-torque location and grew into Jupiter (Cossou et al 2014), carving a radial gap in the disk.  The inward migration of more distant embryos was then blocked by Jupiter (Izidoro et al 2015a).  Uranus and Neptune, and even Saturn's core, may thus have been trapped in the outer Solar System by Jupiter, whereas in absence of Jupiter these planets or their precursors might have conitnued to migrate into the inner solar system becoming close-in super-Earths.  Simulations of this process are indeed the first to successfully reproduce the relative masses of the ice giants (Izidoro et al 2015b).  What remains unclear are the conditions required for the first embryo to grow into Jupiter before other embryos can migrate past. Indeed, simulations of the migration model over-produce systems of super-Earths (e.g., Cossou et al 2014).

Raymond et al (2016) proposed a hybrid pebble+migration model to explain the absence of terrestrial planets closer-in than Mercury.  In their model, Jupiter's core grew quickly in the very inner Solar System, presumably by accreting inward-drifting pebbles.  Following the migration maps of Bitsch et al (2015), the core subsequently migrated back outward to several AU.  During its outward migration, the very inner Solar System was swept clean by resonant shepherding (Tanaka and Ida 1999; Mandell et al 2007).  Simulations show that the sweeping mechanism breaks down at 0.5-1 AU, thus clearing the region interior to Mercury but not the Venus-Earth zone (Raymond et al 2016).  While provocative, this model is admittedly built on several assumptions (pebble drift and outward migration, notably) with associated uncertainties.

Two recent models have proposed that the primordial Solar System contained systems of super-Earths that were later destroyed (Volk \& Gladman 2015; Batygin \& Laughlin 2015). However, each of these studies has significant limitations. For instance, Volk \& Gladman (2015) proposed that the Solar System's super-Earths were destabilized and underwent a series of catastrophic collisions that ground the planets to dust. However, they did not model the actual collisions or collisional grinding, and a cursory examination of the relevant collisional parameters suggests that they were not in the supercatastrophic regime (see Genda et al 2012; Leinhardt \& Stewart 2012). Batygin \& Laughlin (2015) proposed that the inward-then-outward migration proposed by the Grand Tack model (Walsh et al 2011; Pierens \& Raymond 2011) produced a giant pulse of collisional debris that drifted inward and pushed a set of super-Earths onto the Sun. However, Batygin \& Laughlin a) did not consistently treat the accretion of collisional debris in the Earth-Venus zone, b) did not take collisional evolution/growth into account as the debris drifted inward, and c) ignored the fact that disks have inner edges which are inferred from observations (e.g., Eisner et al 2005) and naturally produced in models of gas accretion onto stars (e.g., Romanova et al 2013). Taking these limitations into account, we do not find either of these studies convincing.  (For a more detailed critique of these studies we refer the reader to section 3.1 of Raymond et al (2016).)  Yet while the proposed mechanisms are not compelling, it remains a possibility that the Solar System once contained a system of super-Earths.  What is missing are a viable destruction mechanism and, ideally, an observational signature.

\section{Discussion: is the Solar System special?}
\label{special}

Just like any individual person, the Solar System has its own history. The probability of any other planetary system following an identical blueprint is zero. But how typical was the Solar System's evolutionary path?  {\it (Did it go to college, get married and get a normal job?)}  Or was the Solar System's path unusual in some way?  {\it (Did it quit high school to teach scuba diving in Belize?)}

Jupiter is the only Solar System planet within the reach of current observations.  Based on statistically-sound exoplanet observational surveys, the Sun-Jupiter system is special at roughly the level of one in a thousand.  First, the Sun is an unusually massive star; the most common type of star are M dwarfs, with masses of 10-50\% of the Sun's. The Sun is unusual at the $\sim$10\% level, depending on the definition of ``Sun-like'' and the assumed stellar initial mass function.  Second, only $\sim$10\% of Sun-like stars have gas giant planets with orbits shorter than a few to 10 AU (Cumming et al 2008; Mayor et al 2011). Third, only about 10\% of giant exoplanets have orbits wider than 1 AU and eccentricities smaller than 0.1. The median eccentricity is $\sim$0.25 (Butler et al 2006; Udry \& Santos 2007).  Taken together, these constraints suggest the Sun-Jupiter system is a 0.1\% case.  Of course, the situation is more complex than described here, as there exist correlations between a number of relevant parameters such as between the stellar mass and the presence of gas giants (Johnson et al 2007; Lovis \& Mayor 2007) and between the orbital radius and eccentricity (e.g., Butler et al 2006; Ford \& Rasio 2008). The numbers quoted here are a simple order of magnitude, but they clearly illustrate that the Solar System is not a typical case in at least one regard: the presence and orbit of Jupiter.

Currently-favored theoretical modeling likewise favors Jupiter as the key player in the Solar System's evolution. The Grand Tack model invokes a two phase migration history for Jupiter (Walsh et al 2011).  Izidoro et al (2015a) proposed that Jupiter held back the inward migration of Uranus and Neptune and stopped them from become hot super-Earths. Assorted other models invoke Jupiter's growth (Turrini et al 2012), dynamics (Izidoro et al 2016) or migration (Batygin \& Laughlin 2015) to explain the properties of the Solar System.  

There remain considerable uncertainties at each step of planet formation. That is the theme of this review: even our most successful models are built on a shaky foundation.  Looking to the future, the most important steps forward may be those that show how planet do {\em not} form.

\acknowledgments
We acknowledge support by the French ANR, project number ANR-13--13-BS05-0003-01  projet MOJO (Modeling the Origin of JOvian planets). 
This review paper does not report any new data. All data reported in the figures or in the text are derived from the cited sources. 

\section{References}

\begin{itemize}

\item[--] Adams, E.~R., Seager, S., Elkins-Tanton, L.\ 2008.\ Ocean Planet or Thick Atmosphere: On the Mass-Radius Relationship for Solid Exoplanets with Massive Atmospheres.\ The Astrophysical Journal 673, 1160-1164. 

\item[--] Alexander, R., Pascucci, I., Andrews, S., Armitage, P., Cieza, L.\ 2014.\ The Dispersal of Protoplanetary Disks.\ Protostars and Planets VI 475-496. 

\item[--] Alibert, Y., Mordasini, C., Benz, W., Winisdoerffer, C.\ 2005.\ Models of giant planet formation with migration and disc evolution.\ Astronomy and Astrophysics 434, 343-353.

\item[--] Andrews, S.~M., Wilner, D.~J., Hughes, A.~M., Qi, C., Dullemond, C.~P.\ 2009.\ Protoplanetary Disk Structures in Ophiuchus.\ The Astrophysical Journal 700, 1502-1523. 

\item[--] Armitage, P.~J.\ 2011.\ Dynamics of Protoplanetary Disks.\ Annual Review of Astronomy and Astrophysics 49, 195-236. 

\item[--] Asphaug, E., Jutzi, M., Movshovitz, N.\ 2011.\ Chondrule formation during planetesimal accretion.\ Earth and Planetary Science Letters 308, 369-379. 

\item[--] Bai, X.-N.\ 2013.\ Wind-driven Accretion in Protoplanetary Disks. II. Radial Dependence and Global Picture.\ The Astrophysical Journal 772, 96. 

\item[--] Bai, X.-N., Stone, J.~M.\ 2013.\ Wind-driven Accretion in Protoplanetary Disks. I. Suppression of the Magnetorotational Instability and Launching of the Magnetocentrifugal Wind.\ The Astrophysical Journal 769, 76. 

\item[--] Bai, X.-N.\ 2014.\ Hall-effect-Controlled Gas Dynamics in Protoplanetary Disks. I. Wind Solutions at the Inner Disk.\ The Astrophysical Journal 791, 137. 

\item[--] Bai, X.-N.\ 2015.\ Hall Effect Controlled Gas Dynamics in Protoplanetary Disks. II. Full 3D Simulations toward the Outer Disk.\ The Astrophysical Journal 798, 84. 

\item[--] Bai, X.-N.\ 2016.\ Towards a Global Evolutionary Model of Protoplanetary Disks.\ The Astrophysical Journal 821, 80. 

\item[--] Balbus, S.~A., Hawley, J.~F.\ 1991.\ A powerful local shear instability in weakly magnetized disks. I - Linear analysis. II - Nonlinear evolution.\ The Astrophysical Journal 376, 214-233. 

\item[--] Balbus, S.~A., Papaloizou, J.~C.~B.\ 1999.\ On the Dynamical Foundations of {$\alpha$} Disks.\ The Astrophysical Journal 521, 650-658. 

\item[--] Batygin, K., Laughlin, G.\ 2015.\ Jupiter's decisive role in the inner Solar System's early evolution.\ Proceedings of the National Academy of Science 112, 4214-4217. 

\item[--] Ben{\'{\i}}tez-Llambay, P., Masset, F., Koenigsberger, G., Szul{\'a}gyi, J.\ 2015.\ Planet heating prevents inward migration of planetary cores.\ Nature 520, 63-65. 

\item[--] Benz, W., Slattery, W.~L., Cameron, A.~G.~W.\ 1986.\ The origin of the moon and the single-impact hypothesis. I.\ Icarus 66, 515-535. 

\item[--] Birnstiel, T., Andrews, S.~M.\ 2014.\ On the Outer Edges of Protoplanetary Dust Disks.\ The Astrophysical Journal 780, 153. 

\item[--] Birnstiel, T., Fang, M., Johansen, A.\ 2016.\ Dust Evolution and the Formation of Planetesimals.\ ArXiv e-prints arXiv:1604.02952. 

\item[--] Bitsch, B., Kley, W.\ 2010.\ Orbital evolution of eccentric planets in radiative discs.\ Astronomy and Astrophysics 523, A30. 

\item[--] Bitsch, B., Morbidelli, A., Lega, E., Crida, A.\ 2014.\ Stellar irradiated discs and implications on migration of embedded planets. II. Accreting-discs.\ Astronomy and Astrophysics 564, A135. 

\item[--] Bitsch, B., Morbidelli, A., Lega, E., Kretke, K., Crida, A.\ 2014b.\ Stellar irradiated discs and implications on migration of embedded planets. III. Viscosity transitions.\ Astronomy and Astrophysics 570, A75. 

\item[--] Bitsch, B., Johansen, A., Lambrechts, M., Morbidelli, A.\ 2015.\ The structure of protoplanetary discs around evolving young stars.\ Astronomy and Astrophysics 575, A28. 

\item[--] Bitsch, B., Lambrechts, M., Johansen, A.\ 2015b.\ The growth of planets by pebble accretion in evolving protoplanetary discs.\ Astronomy and Astrophysics 582, A112. 

\item[--] Blum, J., Wurm, G.\ 2008.\ The Growth Mechanisms of Macroscopic Bodies in Protoplanetary Disks.\ Annual Review of Astronomy and Astrophysics 46, 21-56. 

\item[--] Bolmont, E., Raymond, S.~N., von Paris, P., Selsis, F., Hersant, F., Quintana, E.~V., Barclay, T.\ 2014.\ Formation, Tidal Evolution, and Habitability of the Kepler-186 System.\ The Astrophysical Journal 793, 3. 

\item[--] Bottke, W.~F., Nesvorn{\'y}, D., Grimm, R.~E., Morbidelli, A., O'Brien, D.~P.\ 2006.\ Iron meteorites as remnants of planetesimals formed in the terrestrial planet region.\ Nature 439, 821-824. 

\item[--] Bouvier, A., Blichert-Toft, J., Moynier, F., Vervoort, J.D., Albarede, F. 2007. Pb - Pb dating constraints on the accretion and cooling history of chondrites. Geochimica et Cosmochimica Acta 71, 1583-1604. 

\item[--] Bouvier, J., and 10 colleagues 2007.\ Magnetospheric accretion-ejection processes in the classical T Tauri star AA Tauri.\ Astronomy and Astrophysics 463, 1017-1028. 

\item[--] Burkhardt, C., Kleine, T., Bourdon, B., Palme, H., Zipfel, J., Friedrich, J.M., Ebel, D.S. 2008. Hf - W mineral isochron for Ca,Al-rich inclusions: Age of the solar system and the timing of core formation in planetesimals. Geochimica et Cosmochimica Acta 72, 6177-6197. 

\item[--] Butler, R.~P., Johnson, J.~A., Marcy, G.~W., Wright, J.~T., Vogt, S.~S., Fischer, D.~A.\ 2006.\ A Long-Period Jupiter-Mass Planet Orbiting the Nearby M Dwarf GJ 849.\ Publications of the Astronomical Society of the Pacific 118, 1685-1689. 

\item[--] Campante, T.~L., and 40 colleagues 2015.\ An Ancient Extrasolar System with Five Sub-Earth-size Planets.\ The Astrophysical Journal 799, 170. 

\item[--] Canup, R.~M., Asphaug, E.\ 2001.\ Origin of the Moon in a giant impact near the end of the Earth's formation.\ Nature 412, 708-712. 

\item[--] Carrera, D., Johansen, A., Davies, M.~B.\ 2015.\ How to form planetesimals from mm-sized chondrules and chondrule aggregates.\ Astronomy and Astrophysics 579, A43. 

\item[--] Chatterjee, S., Tan, J.~C.\ 2014.\ Inside-out Planet Formation.\ The Astrophysical Journal 780, 53. 

\item[--] Chatterjee, S., Tan, J.~C.\ 2015.\ Vulcan Planets: Inside-out Formation of the Innermost Super-Earths.\ The Astrophysical Journal 798, L32. 

\item[--] Chiang, E.~I., 
Goldreich, P.\ 1997.\ Spectral Energy Distributions of T Tauri Stars with 
Passive Circumstellar Disks.\ The Astrophysical Journal 490, 368-376. 

\item[--] Chiang, E., Youdin, A.~N.\ 2010.\ Forming Planetesimals in Solar and Extrasolar Nebulae.\ Annual Review of Earth and Planetary Sciences 38, 493-522. 

\item[--] Chiang, E., Laughlin, G.\ 2013.\ The minimum-mass extrasolar nebula: in situ formation of close-in super-Earths.\ Monthly Notices of the Royal Astronomical Society 431, 3444-3455. 

\item[--] Coleman, G.~A.~L., Nelson, R.~P.\ 2014.\ On the formation of planetary systems via oligarchic growth in thermally evolving viscous discs.\ Monthly Notices of the Royal Astronomical Society 445, 479-499. 

\item[--] Coleman, G.~A.~L., Nelson, R.~P.\ 2016.\ Giant planet formation in radially structured protoplanetary discs. Monthly Notices of the Royal Astronomical Society 460, 2779-2795.  

\item[--] Cossou, C., Raymond, S.~N., Hersant, F., Pierens, A.\ 2014.\ Hot super-Earths and giant planet cores from different migration histories.\ Astronomy and Astrophysics 569, A56. 

\item[--] Crida, A., Morbidelli, A., Masset, F.\ 2006.\ On the width and shape of gaps in protoplanetary disks.\ Icarus 181, 587-604. 

\item[--] Cumming, A., Butler, R.~P., Marcy, G.~W., Vogt, S.~S., Wright, J.~T., Fischer, D.~A.\ 2008.\ The Keck Planet Search: Detectability and the Minimum Mass and Orbital Period Distribution of Extrasolar Planets.\ Publications of the Astronomical Society of the Pacific 120, 531-554.

\item[--] Cuzzi, J.~N., Hogan, R.~C., Paque, J.~M., Dobrovolskis, A.~R.\ 2001.\ Size-selective Concentration of Chondrules and Other Small Particles in Protoplanetary Nebula Turbulence.\ The Astrophysical Journal 546, 496-508. 

\item[--] Cuzzi, J.~N., Hogan, R.~C., Bottke, W.~F.\ 2010.\ Towards initial mass functions for asteroids and Kuiper Belt Objects.\ Icarus 208, 518-538. 

\item[--] Cuzzi, J.~N., Hartlep, T., Estrada, P.~R.\ 2016.\ Planetesimal Initial Mass Functions and Creation Rates Under Turbulent Concentration Using Scale-Dependent Cascades.\ Lunar and Planetary Science Conference 47, 2661. 

\item[--] Dauphas, N., Pourmand, A.\ 2011.\ Hf-W-Th evidence for rapid growth of Mars and its status as a planetary embryo.\ Nature 473, 489-492. 

\item[--] Deienno, R., Gomes, R.~S., Walsh, K.~J., Morbidelli, A., Nesvorn{\'y}, D.\ 2016.\ Is the Grand Tack model compatible with the orbital distribution of main belt asteroids?.\ Icarus 272, 114-124. 

\item[--] Dominik, C., Tielens, A.~G.~G.~M.\ 1997.\ The Physics of Dust Coagulation and the Structure of Dust Aggregates in Space.\ The Astrophysical Journal 480, 647-673. 

\item[--] Dr{\c a}{\.z}kowska, J., Dullemond, C.~P.\ 2014.\ Can dust coagulation trigger streaming instability?.\ Astronomy and Astrophysics 572, A78. 

\item[--] Dr{\c a}{\.z}kowska, J., Alibert, Y. and Moore, B. 2016. Close-in planetesimal formation by pile-up of drifting pebbles. Astron. and Astrophys, in press.

\item[--] Dressing, C.~D., Charbonneau, D., Newton, E.~R.\ 2015.\ The Occurrence Rate and Composition of Small Planets Orbiting Small Stars.\ AAS/Division for Extreme Solar Systems Abstracts 3, 501.03. 

\item[--] Duffell, P.~C., Haiman, Z., MacFadyen, A.~I., D'Orazio, D.~J., Farris, B.~D.\ 2014.\ The Migration of Gap-opening Planets is Not Locked to Viscous Disk Evolution.\ The Astrophysical Journal 792, L10. 

\item[--] Dullemond, C.~P., 
Dominik, C., Natta, A.\ 2001.\ Passive Irradiated Circumstellar Disks with 
an Inner Hole.\ The Astrophysical Journal 560, 957-969. 

\item[--] Dullemond, C.~P., van Zadelhoff, G.~J., Natta, A.\ 2002.\ Vertical structure models of T Tauri and Herbig Ae/Be disks.\ Astronomy and Astrophysics 389, 464-474. 

\item[--] D{\"u}rmann, C., Kley, W.\ 2015.\ Migration of massive planets in accreting disks.\ Astronomy and Astrophysics 574, A52. 

\item[--] Eisner, J.~A., Hillenbrand, L.~A., White, R.~J., Akeson, R.~L., Sargent, A.~I.\ 2005.\ Observations of T Tauri Disks at Sub-AU Radii: Implications for Magnetospheric Accretion and Planet Formation.\ The Astrophysical Journal 623, 952-966. 

\item[--] Fabrycky, D.~C., and 21 colleagues 2014.\ Architecture of Kepler's Multi-transiting Systems. II. New Investigations with Twice as Many Candidates.\ The Astrophysical Journal 790, 146. 

\item[--] Fang, J., Margot, J.-L.\ 2012.\ Architecture of Planetary Systems Based on Kepler Data: Number of Planets and Coplanarity.\ The Astrophysical Journal 761, 92. 

\item[--] Fang, J., Margot, J.-L.\ 2013.\ Are Planetary Systems Filled to Capacity? A Study Based on Kepler Results.\ The Astrophysical Journal 767, 115. 

\item[--] Ford, E.~B., Rasio, F.~A.\ 2008.\ Origins of Eccentric Extrasolar Planets: Testing the Planet-Planet Scattering Model.\ The Astrophysical Journal 686, 621-636. 

\item[--] Fressin, F., Torres, G., Charbonneau, D., Bryson, S.~T., Christiansen, J., Dressing, C.~D., Jenkins, J.~M., Walkowicz, L.~M., Batalha, N.~M.\ 2013.\ The False Positive Rate of Kepler and the Occurrence of Planets.\ The Astrophysical Journal 766, 81. 

\item[--] Fung, J., Artymowicz, P., Wu, Y.\ 2015.\ The 3D Flow Field Around an Embedded Planet.\ The Astrophysical Journal 811, 101. 

\item[--] Genda, H., Kokubo, E., Ida, S.\ 2012.\ Merging Criteria for Giant Impacts of Protoplanets.\ The Astrophysical Journal 744, 137. 

\item[--] Goldreich, P., Tremaine, S.\ 1980.\ Disk-satellite interactions.\ The Astrophysical Journal 241, 425-441. 

\item[--] Gressel, O., Turner, N.~J., Nelson, R.~P., McNally, C.~P.\ 2015.\ Global Simulations of Protoplanetary Disks With Ohmic Resistivity and Ambipolar Diffusion.\ The Astrophysical Journal 801, 84. 

\item[--] Guillot, T., Ida, S., Ormel, C.~W.\ 2014.\ On the filtering and processing of dust by planetesimals. I. Derivation of collision probabilities for non-drifting planetesimals.\ Astronomy and Astrophysics 572, A72. 

\item[--] G{\"u}ttler, C., 
Blum, J., Zsom, A., Ormel, C.~W., Dullemond, C.~P.\ 2009.\ The first phase 
of protoplanetary dust growth: The bouncing barrier.\ Geochimica et 
Cosmochimica Acta Supplement 73, 482. 

\item[--] Hansen, B.~M.~S.\ 2009.\ Formation of the Terrestrial Planets from a Narrow Annulus.\ The Astrophysical Journal 703, 1131-1140. 

\item[--] Hansen, B.~M.~S., Murray, N.\ 2013.\ Testing in Situ Assembly with the Kepler Planet Candidate Sample.\ The Astrophysical Journal 775, 53. 

\item[--] Haisch, K.~E., Jr., Lada, E.~A., Lada, C.~J.\ 2001.\ Disk Frequencies and Lifetimes in Young Clusters.\ The Astrophysical Journal 553, L153-L156. 

\item[--] Hasegawa, Y., Pudritz, R.~E.\ 2011.\ The origin of planetary system architectures - I. Multiple planet traps in gaseous discs.\ Monthly Notices of the Royal Astronomical Society 417, 1236-1259. 

\item[--] Hellary, P., Nelson, R.~P.\ 2012.\ Global models of planetary system formation in radiatively-inefficient protoplanetary discs.\ Monthly Notices of the Royal Astronomical Society 419, 2737-2757. 

\item[--] Helled, R., Bodenheimer, P., Podolak, M., Boley, A., Meru, F., Nayakshin, S., Fortney, J.~J., Mayer, L., Alibert, Y., Boss, A.~P.\ 2014.\ Giant Planet Formation, Evolution, and Internal Structure.\ Protostars and Planets VI 643-665. 

\item[--] Hillenbrand, L.~A.\ 2008.\ Disk-dispersal and planet-formation timescales.\ Physica Scripta Volume T 130, 014024. 

\item[--] Howard, A.~W., and 66 colleagues 2012.\ Planet Occurrence within 0.25 AU of Solar-type Stars from Kepler.\ The Astrophysical Journal Supplement Series 201, 15. 

\item[--] Hu, X., Zhu, Z., Tan, J.~C., Chatterjee, S.\ 2016.\ Inside-out Planet Formation. III. Planet-Disk Interaction at the Dead Zone Inner Boundary.\ The Astrophysical Journal 816, 19. 

\item[--] Hubickyj, O., Bodenheimer, P., Lissauer, J.~J.\ 2005.\ Accretion of the gaseous envelope of Jupiter around a 5 10 Earth-mass core.\ Icarus 179, 415-431. 

\item[--] Ida, S., Lin, D.~N.~C.\ 2004a.\ Toward a Deterministic Model of Planetary Formation. I. A Desert in the Mass and Semimajor Axis Distributions of Extrasolar Planets.\ The Astrophysical Journal 604, 388-413. 

\item[--] Ida, S., Lin, D.~N.~C.\ 2004b.\ Toward a Deterministic Model of Planetary Formation. II. The Formation and Retention of Gas Giant Planets around Stars with a Range of Metallicities.\ The Astrophysical Journal 616, 567-572. 

\item[--] Ida, S., Lin, D.~N.~C.\ 2010.\ Toward a Deterministic Model of Planetary Formation. VI. Dynamical Interaction and Coagulation of Multiple Rocky Embryos and Super-Earth Systems around Solar-type Stars.\ The Astrophysical Journal 719, 810-830. 

\item[--] Ida, S., Guillot, T., Morbidelli, A.\ 2016.\ The radial dependence of pebble accretion rates: A source of diversity in planetary systems I. Analytical formulation. Astronomy and Astrophysics 591, A72. 

\item[--] Ikoma, M., Emori, H., Nakazawa, K.\ 2001.\ Formation of Giant Planets in Dense Nebulae: Critical Core Mass Revisited.\ The Astrophysical Journal 553, 999-1005. 

\item[--] Inamdar, N.~K., Schlichting, H.~E.\ 2015.\ The formation of super-Earths and mini-Neptunes with giant impacts.\ Monthly Notices of the Royal Astronomical Society 448, 1751-1760. 

\item[--] Inamdar, N.~K., Schlichting, H.~E.\ 2016.\ Stealing the Gas: Giant Impacts and the Large Diversity in Exoplanet Densities.\ The Astrophysical Journal 817, L13. 

\item[--] Izidoro, A., Raymond, S.~N., Morbidelli, A., Hersant, F., Pierens, A.\ 2015a.\ Gas Giant Planets as Dynamical Barriers to Inward-Migrating Super-Earths.\ The Astrophysical Journal 800, L22. 

\item[--] Izidoro, A., Morbidelli, A., Raymond, S.~N., Hersant, F., Pierens, A.\ 2015b.\ Accretion of Uranus and Neptune from inward-migrating planetary embryos blocked by Jupiter and Saturn.\ Astronomy and Astrophysics 582, A99. 

\item[--] Izidoro, A., Raymond, S.~N., Morbidelli, A., Winter, O.~C.\ 2015c.\ Terrestrial planet formation constrained by Mars and the structure of the asteroid belt.\ Monthly Notices of the Royal Astronomical Society 453, 3619-3634. 

\item[--] Izidoro, A., Raymond, S.N., Pierens, A., Morbidelli, A., Winter, O.C., Nesvorny, D., 2016. Structure of the Asteroid Belt from the Gas Giants’ 1
Growth and Chaotic Dynamics. Science Advances, in press. ArXiv e-prints arXiv:1609.04970.


\item[--] Jacobson, S.~A., Morbidelli, A.\ 2014.\ Lunar and terrestrial planet formation in the Grand Tack scenario.\ Philosophical Transactions of the Royal Society of London Series A 372, 0174. 

\item[--] Jacobson, S.~A., Morbidelli, A., Raymond, S.~N., O'Brien, D.~P., Walsh, K.~J., Rubie, D.~C.\ 2014.\ Highly siderophile elements in Earth's mantle as a clock for the Moon-forming impact.\ Nature 508, 84-87. 

\item[--] Johansen, A., Henning, T., Klahr, H.\ 2006.\ Dust Sedimentation and Self-sustained Kelvin-Helmholtz Turbulence in Protoplanetary Disk Midplanes.\ The Astrophysical Journal 643, 1219-1232. 

\item[--] Johansen, A., Oishi, J.~S., Mac Low, M.-M., Klahr, H., Henning, T., Youdin, A.\ 2007.\ Rapid planetesimal formation in turbulent circumstellar disks.\ Nature 448, 1022-1025. 

\item[--] Johansen, A., Youdin, A., Mac Low, M.-M.\ 2009.\ Particle Clumping and Planetesimal Formation Depend Strongly on Metallicity.\ The Astrophysical Journal 704, L75-L79. 

\item[--] Johansen, A., Lacerda, P.\ 2010.\ Prograde rotation of protoplanets by accretion of pebbles in a gaseous environment.\ Monthly Notices of the Royal Astronomical Society 404, 475-485. 

\item[--] Johansen, A., Blum, J., Tanaka, H., Ormel, C., Bizzarro, M., Rickman, H.\ 2014.\ The Multifaceted Planetesimal Formation Process.\ Protostars and Planets VI 547-570. 

\item[--] Johansen, A., Mac Low, M.-M., Lacerda, P., Bizzarro, M.\ 2015.\ Growth of asteroids, planetary embryos, and Kuiper belt objects by chondrule accretion.\ Science Advances 1, 1500109. 

\item[--] Johnson, J.~A., Butler, R.~P., Marcy, G.~W., Fischer, D.~A., Vogt, S.~S., Wright, J.~T., Peek, K.~M.~G.\ 2007.\ A New Planet around an M Dwarf: Revealing a Correlation between Exoplanets and Stellar Mass.\ The Astrophysical Journal 670, 833-840. 

\item[--] Johnson, B.~C., Minton, D.~A., Melosh, H.~J., Zuber, M.~T.\ 2015.\ Impact jetting as the origin of chondrules.\ Nature 517, 339-341. 

\item[--] Jontof-Hutter, D., Lissauer, J.~J., Rowe, J.~F., Fabrycky, D.~C.\ 2014.\ Kepler-79's Low Density Planets.\ The Astrophysical Journal 785, 15. 

\item[--] Kataoka, A., Tanaka, H., Okuzumi, S., Wada, K.\ 2013.\ Fluffy dust forms icy planetesimals by static compression.\ Astronomy and Astrophysics 557, L4. 
\item[--] Kenyon, S.~J., Hartmann, L.\ 1987.\ Spectral energy distributions of T Tauri stars - Disk flaring and limits on accretion.\ The Astrophysical Journal 323, 714-733. 

\item[--] Klahr, H.~H., Bodenheimer, P.\ 2003.\ Turbulence in Accretion Disks: Vorticity Generation and Angular Momentum Transport via the Global Baroclinic Instability.\ The Astrophysical Journal 582, 869-892. 

\item[--] Kleine, T., Touboul, M., Bourdon, B., Nimmo, F., Mezger, K., Palme, H., Jacobsen, S.~B., Yin, Q.-Z., Halliday, A.~N.\ 2009.\ Hf-W chronology of the accretion and early evolution of asteroids and terrestrial planets.\ Geochimica et Cosmochimica Acta 73, 5150-5188. 

\item[--] Kokubo, E., Ida, S.\ 2002.\ Formation of Protoplanet Systems and Diversity of Planetary Systems.\ The Astrophysical Journal 581, 666-680. 

\item[--] Kretke, K.~A., Lin, D.~N.~C.\ 2007.\ Grain Retention and Formation of Planetesimals near the Snow Line in MRI-driven Turbulent Protoplanetary Disks.\ The Astrophysical Journal 664, L55-L58.

\item[--] Kruijer, T.~S., Sprung, P., Kleine, T., Leya, I., Burkhardt, C., Wieler, R.\ 2012.\ Hf-W chronometry of core formation in planetesimals inferred from weakly irradiated iron meteorites.\ Geochimica et Cosmochimica Acta 99, 287-304. 

\item[--] Kuchner, M.~J.\ 2004.\ A Minimum-Mass Extrasolar Nebula.\ The Astrophysical Journal 612, 1147-1151. 

\item[--] Lambrechts, M., Johansen, A.\ 2012.\ Rapid growth of gas-giant cores by pebble accretion.\ Astronomy and Astrophysics 544, A32. 

\item[--] Lambrechts, M., Johansen, A.\ 2014.\ Forming the cores of giant planets from the radial pebble flux in protoplanetary discs.\ Astronomy and Astrophysics 572, A107. 

\item[--] Laskar, J.\ 1997.\ Large scale chaos and the spacing of the inner planets..\ Astronomy and Astrophysics 317, L75-L78. 

\item[--] Laughlin, G., Adams, F.~C.\ 1997.\ Possible Stellar Metallicity Enhancements from the Accretion of Planets.\ The Astrophysical Journal 491, L51-L54. 

\item[--] Lee, E.~J., Chiang, E., Ormel, C.~W.\ 2014.\ Make Super-Earths, Not Jupiters: Accreting Nebular Gas onto Solid Cores at 0.1 AU and Beyond.\ The Astrophysical Journal 797, 95. 

\item[--] Leinhardt, Z.~M., Stewart, S.~T.\ 2012.\ Collisions between Gravity-dominated Bodies. I. Outcome Regimes and Scaling Laws.\ The Astrophysical Journal 745, 79. 

\item[--] Lesur, G., Kunz, M.~W., Fromang, S.\ 2014.\ Thanatology in protoplanetary discs. The combined influence of Ohmic, Hall, and ambipolar diffusion on dead zones.\ Astronomy and Astrophysics 566, A56. 

\item[--] Levison, H.~F., Bottke, W.~F., Gounelle, M., Morbidelli, A., Nesvorn{\'y}, D., Tsiganis, K.\ 2009.\ Contamination of the asteroid belt by primordial trans-Neptunian objects.\ Nature 460, 364-366. 

\item[--] Levison, H.~F., Kretke, K.~A., Duncan, M.~J.\ 2015.\ Growing the gas-giant planets by the gradual accumulation of pebbles.\ Nature 524, 322-324. 

\item[--] Levison, H.~F., Kretke, K.~A., Walsh, K.~J., Bottke, W.~F.\ 2015b.\ Growing the terrestrial planets from the gradual accumulation of sub-meter sized objects.\ Proceedings of the National Academy of Science 112, 14180-14185. 

\item[--] Libourel, G., Krot, A.~N., Chaussidon, M.\ 2006.\ Olivines in Magnesian Porphyritic Chondrules: Mantle Material of Earlier Generations of Differentiated Planetesimals?.\ Meteoritics and Planetary Science Supplement 41, 5295. 

\item[--] Lin, D.~N.~C., Papaloizou, J.\ 1986.\ On the tidal interaction between protoplanets and the protoplanetary disk. III - Orbital migration of protoplanets.\ The Astrophysical Journal 309, 846-857. 

\item[--] Lin, D.~N.~C., Bodenheimer, P., Richardson, D.~C.\ 1996.\ Orbital migration of the planetary companion of 51 Pegasi to its present location.\ Nature 380, 606-607. 

\item[--] Lin, D.~N.~C.\ 1997.\ Planetary Formation in Protostellar Disks.\ IAU Colloq.~163: Accretion Phenomena and Related Outflows 121, 321. 

\item[--] Lissauer, J.~J.\ 2007.\ Planets Formed in Habitable Zones of M Dwarf Stars Probably Are Deficient in Volatiles.\ The Astrophysical Journal 660, L149-L152.

\item[--] Lissauer, J.~J., and 24 colleagues 2011.\ Architecture and Dynamics of Kepler's Candidate Multiple Transiting Planet Systems.\ The Astrophysical Journal Supplement Series 197, 8.

\item[--] Lodders, K.\ 2003.\ Solar 
System Abundances and Condensation Temperatures of the Elements.\ The 
Astrophysical Journal 591, 1220-1247. 

\item[--] Lopez, E.~D., Fortney, J.~J.\ 2014.\ Understanding the Mass-Radius Relation for Sub-neptunes: Radius as a Proxy for Composition.\ The Astrophysical Journal 792, 1. 

\item[--] Lovis, C., Mayor, M.\ 2007.\ Planets around evolved intermediate-mass stars. I. Two substellar companions in the open clusters NGC 2423 and NGC 4349.\ Astronomy and Astrophysics 472, 657-664. 

\item[--] Lynden-Bell, D., Pringle, J.~E.\ 1974.\ The evolution of viscous discs and the origin of the nebular variables..\ Monthly Notices of the Royal Astronomical Society 168, 603-637. 

\item[--] Mandell, A.~M., Raymond, S.~N., Sigurdsson, S.\ 2007.\ Formation of Earth-like Planets During and After Giant Planet Migration.\ The Astrophysical Journal 660, 823-844. 

\item[--] Marcus, P.~S., Pei, S., Jiang, C.-H., Barranco, J.~A., Hassanzadeh, P., Lecoanet, D.\ 2015.\ Zombie Vortex Instability. I. A Purely Hydrodynamic Instability to Resurrect the Dead Zones of Protoplanetary Disks.\ The Astrophysical Journal 808, 87. 

\item[--] Marcy, G.~W., and 102 colleagues 2014.\ Masses, Radii, and Orbits of Small Kepler Planets: The Transition from Gaseous to Rocky Planets.\ The Astrophysical Journal Supplement Series 210, 20. 

\item[--] Martin, R.~G., Lubow, S.~H., Livio, M., Pringle, J.~E.\ 2012.\ Dead zones around young stellar objects: FU Orionis outbursts and transition discs.\ Monthly Notices of the Royal Astronomical Society 423, 2718-2725. 

\item[--] Masset, F., Snellgrove, M.\ 2001.\ Reversing type II migration: resonance trapping of a lighter giant protoplanet.\ Monthly Notices of the Royal Astronomical Society 320, L55-L59. 

\item[--] Masset, F.~S., Morbidelli, A., Crida, A., Ferreira, J.\ 2006.\ Disk Surface Density Transitions as Protoplanet Traps.\ The Astrophysical Journal 642, 478-487. 

\item[--] Matsumoto, Y., Nagasawa, M., Ida, S.\ 2012.\ The orbital stability of planets trapped in the first-order mean-motion resonances.\ Icarus 221, 624-631. 

\item[--] Mayor, M., Queloz, D.\ 1995.\ A Jupiter-mass companion to a solar-type star.\ Nature 378, 355-359. 

\item[--] Mayor, M., and 13 colleagues 2011.\ The HARPS search for southern extra-solar planets XXXIV. Occurrence, mass distribution and orbital properties of super-Earths and Neptune-mass planets.\ ArXiv e-prints arXiv:1109.2497. 

\item[--] McNeil, D.~S., Nelson, R.~P.\ 2010.\ On the formation of hot Neptunes and super-Earths.\ Monthly Notices of the Royal Astronomical Society 401, 1691-1708. 

\item[--] Mills, S., Fabrycky, D.~C., Migaszewski, C., Ford, E.~B., Petigura, E., Isaacson, H.~T.\ 2016.\ Kepler-223: A Resonant Chain of Four Transiting, Sub-Neptune Planets.\ AAS/Division of Dynamical Astronomy Meeting 47, \#101.03. 

\item[--] Morbidelli, A., Crida, A.\ 2007.\ The dynamics of Jupiter and Saturn in the gaseous protoplanetary disk.\ Icarus 191, 158-171. 

\item[--] Morbidelli, A., Lunine, J.~I., O'Brien, D.~P., Raymond, S.~N., Walsh, K.~J.\ 2012.\ Building Terrestrial Planets.\ Annual Review of Earth and Planetary Sciences 40, 251-275. 

\item[--] Morbidelli, A., Szul{\'a}gyi, J., Crida, A., Lega, E., Bitsch, B., Tanigawa, T., Kanagawa, K.\ 2014.\ Meridional circulation of gas into gaps opened by giant planets in three-dimensional low-viscosity disks.\ Icarus 232, 266-270. 

\item[--] Morbidelli, A., Lambrechts, M., Jacobson, S., Bitsch, B.\ 2015.\ The great dichotomy of the Solar System: Small terrestrial embryos and massive giant planet cores.\ Icarus 258, 418-429.

\item[--] Morbidelli, A., Bitsch, B., Crida, A., Gounelle, M., Guillot, T., Jacobson, S., Johansen, A., Lambrechts, M., Lega, E.\ 2016.\ Fossilized condensation lines in the Solar System protoplanetary disk.\ Icarus 267, 368-376. 

\item[--] Moriarty, J., Fischer, D.\ 2015.\ Building Massive Compact Planetesimal Disks from the Accretion of Pebbles.\ The Astrophysical Journal 809, 94. 

\item[--] Mulders, G.~D., Pascucci, I., Apai, D.\ 2015.\ An Increase in the Mass of Planetary Systems around Lower-mass Stars.\ The Astrophysical Journal 814, 130. 

\item[--] Nelson, R.~P., Gressel, O., Umurhan, O.~M.\ 2013.\ Linear and non-linear evolution of the vertical shear instability in accretion discs.\ Monthly Notices of the Royal Astronomical Society 435, 2610-2632. 

\item[--] O'Brien, D.~P., Morbidelli, A., Levison, H.~F.\ 2006.\ Terrestrial planet formation with strong dynamical friction.\ Icarus 184, 39-58. 

\item[--] Ogihara, M., Ida, S.\ 2009.\ N-Body Simulations of Planetary Accretion Around M Dwarf Stars.\ The Astrophysical Journal 699, 824-838. 

\item[--] Ogihara, M., Morbidelli, A., Guillot, T.\ 2015a.\ A reassessment of the in situ formation of close-in super-Earths.\ Astronomy and Astrophysics 578, A36. 

\item[--] Ogihara, M., Morbidelli, A., Guillot, T.\ 2015b.\ Suppression of type I migration by disk winds.\ Astronomy and Astrophysics 584, L1. 

\item[--] Okuzumi, S., Tanaka, H., Kobayashi, H., Wada, K.\ 2012.\ Rapid Coagulation of Porous Dust Aggregates outside the Snow Line: A Pathway to Successful Icy Planetesimal Formation.\ The Astrophysical Journal 752, 106. 

\item[--] Ogilvie, G.~I., Lubow, S.~H.\ 2002.\ On the wake generated by a planet in a disc.\ Monthly Notices of the Royal Astronomical Society 330, 950-954. 

\item[--] Ormel, C.~W., Klahr, H.~H.\ 2010.\ The effect of gas drag on the growth of protoplanets. Analytical expressions for the accretion of small bodies in laminar disks.\ Astronomy and Astrophysics 520, A43. 

\item[--] Paardekooper, S.-J., Mellema, G.\ 2006.\ Halting type I planet migration in non-isothermal disks.\ Astronomy and Astrophysics 459, L17-L20. 

\item[--] Paardekooper, S.-J., Baruteau, C., Crida, A., Kley, W.\ 2010.\ A torque formula for non-isothermal type I planetary migration - I. Unsaturated horseshoe drag.\ Monthly Notices of the Royal Astronomical Society 401, 1950-1964. 

\item[--] Paardekooper, S.-J., Baruteau, C., Kley, W.\ 2011.\ A torque formula for non-isothermal Type I planetary migration - II. Effects of diffusion.\ Monthly Notices of the Royal Astronomical Society 410, 293-303. 

\item[--] Petigura, E.~A., Marcy, G.~W., Howard, A.~W.\ 2013.\ A Plateau in the Planet Population below Twice the Size of Earth.\ The Astrophysical Journal 770, 69. 

\item[--] Pfalzner, S., Steinhausen, M., Menten, K.\ 2014.\ Short Dissipation Times of Proto-planetary Disks: An Artifact of Selection Effects?.\ The Astrophysical Journal 793, L34.

\item[--] Pierens, A., Nelson, R.~P.\ 2008.\ Constraints on resonant-trapping for two planets embedded in a protoplanetary disc.\ Astronomy and Astrophysics 482, 333-340. 

\item[--] Pierens, A., Raymond, S.~N.\ 2011.\ Two phase, inward-then-outward migration of Jupiter and Saturn in the gaseous solar nebula.\ Astronomy and Astrophysics 533, A131. 

\item[--] Pollack, J.~B., Hubickyj, O., Bodenheimer, P., Lissauer, J.~J., Podolak, M., Greenzweig, Y.\ 1996.\ Formation of the Giant Planets by Concurrent Accretion of Solids and Gas.\ Icarus 124, 62-85. 

\item[--] Pringle, J.~E.\ 1981.\ Accretion discs in astrophysics.\ Annual Review of Astronomy and Astrophysics 19, 137-162. 

\item[--] Pu, B., Wu, Y.\ 2015.\ Spacing of Kepler Planets: Sculpting by Dynamical Instability.\ The Astrophysical Journal 807, 44. 

\item[--] Raymond, S.~N., Quinn, T., Lunine, J.~I.\ 2006.\ High-resolution simulations of the final assembly of Earth-like planets I. Terrestrial accretion and dynamics.\ Icarus 183, 265-282. 

\item[--] Raymond, S.\ 2007.\ Formation and Stability of ''Hot Earth'' Planets.\ Bulletin of the American Astronomical Society 39, 110.03. 

\item[--] Raymond, S.~N., Scalo, J., Meadows, V.~S.\ 2007.\ A Decreased Probability of Habitable Planet Formation around Low-Mass Stars.\ The Astrophysical Journal 669, 606-614.

\item[--] Raymond, S.~N., Barnes, R., Mandell, A.~M.\ 2008.\ Observable consequences of planet formation models in systems with close-in terrestrial planets.\ Monthly Notices of the Royal Astronomical Society 384, 663-674. 

\item[--] Raymond, S.~N., O'Brien, D.~P., Morbidelli, A., Kaib, N.~A.\ 2009.\ Building the terrestrial planets: Constrained accretion in the inner Solar System.\ Icarus 203, 644-662. 

\item[--] Raymond, S.~N., Kokubo, E., Morbidelli, A., Morishima, R., Walsh, K.~J.\ 2014.\ Terrestrial Planet Formation at Home and Abroad.\ Protostars and Planets VI 595-618. 

\item[--] Raymond, S.~N., Morbidelli, A.\ 2014.\ The Grand Tack model: a critical review.\ IAU Symposium 310, 194-203. 

\item[--] Raymond, S.~N., Cossou, C.\ 2014.\ No universal minimum-mass extrasolar nebula: evidence against in situ accretion of systems of hot super-Earths.\ Monthly Notices of the Royal Astronomical Society 440, L11-L15. 

\item[--] Raymond, S.~N., Izidoro, A., Bitsch, B., Jacobson, S.~A.\ 2016.\ Did Jupiter's core form in the innermost parts of the Sun's protoplanetary disc?.\ Monthly Notices of the Royal Astronomical Society 458, 2962-2972. 

\item[--] Rogers, L.~A., Bodenheimer, P., Lissauer, J.~J., Seager, S.\ 2011.\ Formation and Structure of Low-density exo-Neptunes.\ The Astrophysical Journal 738, 59. 

\item[--] Rogers, L.~A.\ 2015.\ Most 1.6 Earth-radius Planets are Not Rocky.\ The Astrophysical Journal 801, 41. 

\item[--] Romanova, M.~M., Ustyugova, G.~V., Koldoba, A.~V., Lovelace, R.~V.~E.\ 2013.\ Warps, bending and density waves excited by rotating magnetized stars: results of global 3D MHD simulations.\ Monthly Notices of the Royal Astronomical Society 430, 699-724. 

\item[--] Saito, E., Sirono, S.-i.\ 2011.\ Planetesimal Formation by Sublimation.\ The Astrophysical Journal 728, 20. 

\item[--] Sato, T., Okuzumi, S., Ida, S.\ 2016.\ On the water delivery to terrestrial embryos by ice pebble accretion.\ Astronomy and Astrophysics 589, A15. 

\item[--] Schlichting, H.\ 2015.\ Orbital Architecture and Mean Motion Resonances of Multiple Planet Systems.\ IAU General Assembly 22, 2255741. 

\item[--] Shakura, N.~I., Sunyaev, R.~A.\ 1973.\ Black holes in binary systems. Observational appearance..\ Astronomy and Astrophysics 24, 337-355. 

\item[--] Simon, J.~B., Lesur, G., Kunz, M.~W., Armitage, P.~J.\ 2015.\ Magnetically driven accretion in protoplanetary discs.\ Monthly Notices of the Royal Astronomical Society 454, 1117-1131. 

\item[--] Stoll, M.~H.~R., Kley, W.\ 2014.\ Vertical shear instability in accretion disc models with radiation transport.\ Astronomy and Astrophysics 572, A77. 

\item[--] Stoll, M.~H.~R., Kley, W.\ 2016.\ Particle dynamics in discs with turbulence generated by the vertical
shear instability. ArXiv e-prints arXiv:1607.02322. 

\item[--] Stone, J.~M., Ostriker, E.~C., Gammie, C.~F.\ 1998.\ Dissipation in Compressible Magnetohydrodynamic Turbulence.\ The Astrophysical Journal 508, L99-L102. 

\item[--] Suzuki, T.~K., Muto, T., Inutsuka, S.-i.\ 2010.\ Protoplanetary Disk Winds via Magnetorotational Instability: Formation of an Inner Hole and a Crucial Assist for Planet Formation.\ The Astrophysical Journal 718, 1289-1304. 

\item[--] Suzuki, T.~K., Ogihara, M., Morbidelli, A., Crida, A. and Guillot, T., 2016. Evolution of Protoplanetary Discs
with Magnetically Driven Disc Winds, Astron. Astrophys., in press

\item[--] Swift, J.~J., Johnson, J.~A., Morton, T.~D., Crepp, J.~R., Montet, B.~T., Fabrycky, D.~C., Muirhead, P.~S.\ 2013.\ Characterizing the Cool KOIs. IV. Kepler-32 as a Prototype for the Formation of Compact Planetary Systems throughout the Galaxy.\ The Astrophysical Journal 764, 105. 

\item[--] Tanaka, H., Ida, S.\ 1999.\ Growth of a Migrating Protoplanet.\ Icarus 139, 350-366. 

\item[--] Tanaka, H., Takeuchi, T., Ward, W.~R.\ 2002.\ Three-Dimensional Interaction between a Planet and an Isothermal Gaseous Disk. I. Corotation and Lindblad Torques and Planet Migration.\ The Astrophysical Journal 565, 1257-1274. 

\item[--] Terquem, C., Papaloizou, J.~C.~B.\ 2007.\ Migration and the Formation of Systems of Hot Super-Earths and Neptunes.\ The Astrophysical Journal 654, 1110-1120. 

\item[--] Throop, H.~B., Bally, J.\ 2005.\ Can Photoevaporation Trigger Planetesimal Formation?.\ The Astrophysical Journal 623, L149-L152. 

\item[--] Tremaine, S., Dong, S.\ 2012.\ The Statistics of Multi-planet Systems.\ The Astronomical Journal 143, 94. 

\item[--] Turner, N.~J., Fromang, S., Gammie, C., Klahr, H., Lesur, G., Wardle, M., Bai, X.-N.\ 2014.\ Transport and Accretion in Planet-Forming Disks.\ Protostars and Planets VI 411-432. 

\item[--] Turrini, D., Coradini, A., Magni, G.\ 2012.\ Jovian Early Bombardment: Planetesimal Erosion in the Inner Asteroid Belt.\ The Astrophysical Journal 750, 8. 

\item[--] Udry, S., Santos, N.~C.\ 2007.\ Statistical Properties of Exoplanets.\ Annual Review of Astronomy and Astrophysics 45, 397-439. 

\item[--] Villeneuve, J., Chaussidon, M., Libourel, G.\ 2009.\ Homogeneous Distribution of $^{26}$Al in the Solar System from the Mg Isotopic Composition of Chondrules.\ Science 325, 985. 

\item[--] Visser, R.~G., Ormel, C.~W.\ 2016.\ On the growth of pebble-accreting planetesimals.\ Astronomy and Astrophysics 586, A66. 

\item[--] Volk, K., Gladman, B.\ 2015.\ Consolidating and Crushing Exoplanets: Did It Happen Here?.\ The Astrophysical Journal 806, L26. 

\item[--] Youdin, A.~N., Goodman, J.\ 2005.\ Streaming Instabilities in Protoplanetary Disks.\ The Astrophysical Journal 620, 459-469. 

\item[--] Walsh, K.~J., Morbidelli, A., Raymond, S.~N., O'Brien, D.~P., Mandell, A.~M.\ 2011.\ A low mass for Mars from Jupiter's early gas-driven migration.\ Nature 475, 206-209. 


\item[--] Ward, W.~R.\ 1997.\ Protoplanet Migration by Nebula Tides.\ Icarus 126, 261-281. 

\item[--] Weidenschilling, S.~J.\ 1977.\ Aerodynamics of solid bodies in the solar nebula.\ Monthly Notices of the Royal Astronomical Society 180, 57-70. 

\item[--] Weidenschilling, S.~J., Cuzzi, J.~N.\ 1993.\ Formation of planetesimals in the solar nebula.\ Protostars and Planets III 1031-1060. 

\item[--] Weidenschilling, S.~J.\ 1995.\ Can Gravitational Instability Form Planetesimals?.\ Icarus 116, 433-435. 

\item[--] Weiss, L.~M., Marcy, G.~W.\ 2014.\ The Mass-Radius Relation for 65 Exoplanets Smaller than 4 Earth Radii.\ The Astrophysical Journal 783, L6. 

\item[--] Wetherill, G.~W.\ 1990.\ Formation of the earth.\ Annual Review of Earth and Planetary Sciences 18, 205-256. 

\item[--] Williams, J.~P., Cieza, L.~A.\ 2011.\ Protoplanetary Disks and Their Evolution.\ Annual Review of Astronomy and Astrophysics 49, 67-117. 

\item[--] Wolfgang, A., Lopez, E.\ 2015.\ How Rocky Are They? The Composition Distribution of Kepler's Sub-Neptune Planet Candidates within 0.15 AU.\ The Astrophysical Journal 806, 183. 

\item[--] Zsom, A., Dullemond, C.~P.\ 2008.\ A representative particle approach to coagulation and fragmentation of dust aggregates and fluid droplets.\ Astronomy and Astrophysics 489, 931-941. 

\end{itemize}

\end{document}